\documentclass[aip,pof,reprint]{revtex4-1}

\usepackage{graphicx}% Include figure files
\begin{document}

%\begin{CJK*}{GB}{gbsn} % Use default fonts from CJK (see below)

% Use the \preprint command to place your local institutional report number
% on the title page in preprint mode.
% Multiple \preprint commands are allowed.
%\preprint{}

\title{Temporal evolution and scaling of mixing in two-dimensional Rayleigh-Taylor turbulence} %Title of paper

% repeat the \author .. \affiliation  etc. as needed
% \email, \thanks, \homepage, \altaffiliation all apply to the current author.
% Explanatory text should go in the []'s,
% actual e-mail address or url should go in the {}'s for \email and \homepage.
% Please use the appropriate macro for the type of information

% \affiliation command applies to all authors since the last \affiliation command.
% The \affiliation command should follow the other information.

\author{Quan ZHOU}
\email[]{Author to whom correspondence should be addressed. Electronic addresses: qzhou@shu.edu.cn}
%\homepage[]{Your web page}
%\thanks{} % Use this in place of foot notes on the title page.
%\altaffiliation{}
\affiliation{$^1$Shanghai Institute of Applied Mathematics and Mechanics, and Shanghai Key Laboratory of Mechanics in Energy Engineering, Shanghai University, Shanghai 200072, China}

% Collaboration name, if desired (requires use of superscriptaddress option in \documentclass).
% \noaffiliation is required (may also be used with the \author command).
%\collaboration{}
%\noaffiliation

\date{\today}

\begin{abstract}
We report a high-resolution numerical study of two-dimensional (2D) miscible Rayleigh-Taylor (RT) incompressible turbulence with the Boussinesq approximation. An ensemble of 100 independent realizations were performed at small Atwood number and unit Prandtl number with a spatial resolution of $2048\times8193$ grid points. Our main focus is on the temporal evolution and the scaling behavior of global quantities and of small-scale turbulence properties. Our results show that the buoyancy force balances the inertial force at all scales below the integral length scale and thus validate the basic force-balance assumption of the Bolgiano-Obukhov scenario in 2D RT turbulence. It is further found that the Kolmogorov dissipation scale $\eta(t)\sim t^{1/8}$, the kinetic-energy dissipation rate $\varepsilon_u(t)\sim t^{-1/2}$, and the thermal dissipation rate $\varepsilon_{\theta}(t)\sim t^{-1}$. All of these scaling properties are in excellent agreement with the theoretical predictions of the Chertkov model [Phys. Rev. Lett. \textbf{91}, 115001 (2003)]. We further discuss the emergence of intermittency and anomalous scaling for high order moments of velocity and temperature differences. The scaling exponents $\xi^r_p$ of the $p$th-order temperature structure functions are shown to saturate to $\xi^r_{\infty}\simeq0.78\pm0.15$ for the highest orders, $p\sim10$. The value of $\xi^r_{\infty}$ and the order at which saturation occurs are compatible with those of turbulent Rayleigh-B\'{e}nard (RB) convection [Phys. Rev. Lett. \textbf{88}, 054503 (2002)], supporting the scenario of universality of buoyancy-driven turbulence with respect to the different boundary conditions characterizing the RT and RB systems.
\end{abstract}

%\keywords{}%Use showkeys class option if keyword display desired

\maketitle %\maketitle must follow title, authors, abstract and \pacs
%\end{CJK*}

% Body of paper goes here. Use proper sectioning commands.
% References should be done using the \cite, \ref, and \label commands
\section{Introduction}

%with the heavy layer being on top of the light layer

Turbulent mixing originated at the interface between two layers of fluids of different densities in a gravitational field, i.e. Rayleigh-Taylor (RT) instability \cite{rayleigh1883prms, tayor1950prsla}, is ubiquitous in nature and in many engineering applications. One can find it in heating of solar coronal \cite{ims2005nature}, in buoyancy-driven mixing in the atmosphere and oceans, in cloud formation \cite{sks2006jas}, and in inertial confinement fusion \cite{twc2002science}. In addition, RT turbulence has been pointed to as the dominant acceleration mechanism for thermonuclear flames in type-Ia supernovae \cite{zwr2005aj, cc2006np}. Other examples of RT instability can also be found in rotational fluids \cite{zhou2002jfm, tao2013pre}. Although RT turbulence is of great importance and has been studied for many decades, there are still some open issues \cite{dyd2004pof, abarzhi2010pttsa}. Specifically, for the past decade, many studies \cite{youngs1999jfm, zhou2001pof, wa2002pof, chertkov2003prl, celani2006prl, matsumoto2009pre, boffetta2009pre, chertkov2009pof, biferale2010pof, boffetta2010pof, abarzhi2010epl, pullin2010jfm, soulard2012pof, soulard2012prl} have focused on small-scale turbulent fluctuations in both two- (2D) and three-dimensional (3D) RT turbulence. In two dimensions, Chertkov \cite{chertkov2003prl} proposed a phenomenological theory. By assuming equipartition of the buoyancy and inertial forces at all scales in the inertial subrange for the energy equation, the model predicts a Bolgiano-Obukhov-like (BO59) \cite{lx2010arfm} scaling for the cascades of both the velocity and temperature fields. This prediction was later confirmed from the view of structure functions by pioneering numerical simulations \cite{celani2006prl, biferale2010pof} and was retrieved from the Monin-Yaglom relation by recent theoretical works \cite{soulard2012prl}.

In this paper, we want to deepen the previous studies \cite{celani2006prl, biferale2010pof} by making a numerical simulation of RT turbulence in the 2D space. Our objective is to study the time evolution and the scaling behavior of the global quantities and of small-scale turbulence properties. Two considerations prompted us to focus on a 2D geometry. First, the numerical effort for 2D simulations is much smaller so that a good resolution becomes feasible for high Reynolds/Rayleigh numbers. Second, in 2D RT turbulence, temperature behaves as an active scalar, leading to the emergence of a BO59 scaling. This is at clear variance with the 3D cases, where temperature is regarded as a passive scalar and thus the Kolmogorov-like (K41) phenomenology was predicted \cite{chertkov2003prl, soulard2012pof, soulard2012prl} and observed \cite{matsumoto2009pre, boffetta2009pre, boffetta2010pof}. There is the attraction of studying the scaling properties in a turbulent system where a non-K41 phenomenology is expected. In addition, the BO59 scaling has long been believed to characterize the cascades of the velocity and temperature fluctuations in turbulent Rayleigh-B\'{e}nard (RB) convection. Despite many years of experimental and numerical investigations, whether the BO59 scaling exists in a turbulent RB system remains unsettled. For more detailed elucidation of the problem, we refer interested readers to the recent review paper by Lohse $\&$ Xia \cite{lx2010arfm}.

The remainder of this paper is organized as follows. In Sec. II we formulate the RT problem and provide the theoretical background. Section III describes the details of the numerical simulations. The numerical results are presented and analyzed in Sec. IV, which is divided into three parts. In Sec. IV A we discuss the statistics of global quantities, Sec. IV B is devoted to the investigation of small-scale properties, and Sec. IV C studies the emergence of intermittency and anomalous scaling for higher-order moments of velocity and temperature differences. Here, we mainly focus on the temporal evolution and scaling. We summarize our findings and conclude in Sec. V.

\section{Rayleigh-Taylor turbulence}

We consider the 2D time-dependent viscous Oberbeck-Boussinesq equations of miscible RT turbulence, namely,
\begin{equation}
\frac{\partial \mathbf{u}}{\partial t}+(\mathbf{u}\cdot\nabla)\mathbf{u}=-\frac{1}{\rho_0}\nabla P+\nu\nabla^2\mathbf{u}+\beta g\theta \vec{z},
\label{eq:ns_v}
\end{equation}
\begin{equation}
\frac{\partial\theta}{\partial t}+(\mathbf{u}\cdot\nabla)\theta=\kappa\nabla^2\theta,
\label{eq:ns_t}
\end{equation}
together with the incompressible condition $\nabla\cdot\mathbf{u}=0$. Here, $\theta(x,z,t)$ is the temperature field, $\mathbf{u}(x,z,t)=u\vec{x}+w\vec{z}$ is the velocity field ($\vec{x}$ and $\vec{z}$ are the horizontal and vertical unit vectors, respectively), $P(x,z,t)$ is the pressure field, $g$ is the acceleration due to gravity, $\beta$, $\nu$, and $\kappa$ are the thermal expansion coefficient, the kinematic viscosity, and the thermal diffusivity of the working fluids, respectively. We assume that $\beta$, $\nu$, and $\kappa$ are the same for the top ($z>0$) and bottom ($z<0$) fluids. In the Oberbeck-Boussinesq approximation, the fluid density $\rho$ is assumed to depend linearly on the temperature, i.e. $\rho=\rho_0[1-\beta(\theta-\theta_0)]$, with $\rho_0$ and $\theta_0$ being reference values.

At the beginning (time $t=0$), the system is at rest ($\mathbf{u}=0$) with the colder fluid being on top of the hotter one. This corresponds to a step function for the initial temperature profile:
\begin{equation}
\theta(x,z,t=0)=-\mathrm{sgn}(z)\Theta_0/2,
\label{eq:itp}
\end{equation}
where $\Theta_0$ is the initial temperature jump which defines the Atwood number as $A=\beta\Theta_0/2$. This initial configuration is unstable and the evolution of the instability results in a mixing zone of the width $h(t)$. Using dimensional analysis and self-similar assumptions \cite{dyd2004pof, clark2004jfm, chertkov2009pof}, one expects that the growth of $h(t)$ follows the accelerated law, i.e.
\begin{equation}
h(t)=\alpha Agt^2,
\label{eq:ht}
\end{equation}
where $\alpha$ is a dimensionless constant the value of which has been studied extensively \cite{dyd2004pof}. The integral length scale $L(t)$ of turbulent flow, defined as the characteristic scale of the production of turbulence, is expected to be linearly related to the geometrical scale $h(t)$,
\begin{equation}
L(t)\sim h(t)\sim\beta g\Theta_0t^2,
\label{eq:lt}
\end{equation}
as shown by recent numerical simulations \cite{chertkov2009pof, boffetta2010pof} for the 3D case. Relation (\ref{eq:lt}) further implies
\begin{equation}
v_{rms}(t)\sim\frac{L(t)}{t}\sim\beta g\Theta_0t
\label{eq:urms}
\end{equation}
for typical velocity fluctuations at the pumping scale, where $v$ denotes one component of the velocity. Usually, relations (\ref{eq:ht})$\sim$(\ref{eq:urms}) are adopted as the central assumptions for some phenomenological models, such as the one advanced by Chertkov \cite{chertkov2003prl}.

In the following, we briefly introduce the main points of the Chertkov model for the 2D case. It is assumed that the buoyancy term on the right-hand side of Eq. (\ref{eq:ns_v}) balances the nonlinear term on the left-hand side of Eq. (\ref{eq:ns_v}) for all scales smaller than the integral one $L(t)$, i.e.,
\begin{equation}
\frac{\delta v_r^2}{r}\sim\beta g\delta\theta_r,
\label{eq:force_balance}
\end{equation}
where $\delta v_r$ and $\delta\theta_r$ are typical velocity and temperature fluctuations at scale $r$, respectively. From this balance, together with the thermal balance from Eq. (\ref{eq:ns_t}),
\begin{equation}
\varepsilon_{\theta}(t)\sim\frac{\delta v_r\delta\theta_r^2}{r}\sim\frac{v_{rms}\Theta_0^2}{L(t)}\sim\frac{\Theta_0^2}{t},
\label{eq:et}
\end{equation}
one immediately arrives at the BO59 scaling,
\begin{equation}
\delta v_r\sim(\frac{r}{L(t)})^{3/5}v_{rms}(t)\sim\frac{r^{3/5}(\beta g\Theta_0)^{2/5}}{t^{1/5}},
\label{eq:sfu}
\end{equation}
\begin{equation}
\delta\theta_r\sim(\frac{r}{L(t)})^{1/5}\Theta_0\sim\frac{r^{1/5}\Theta_0^{4/5}}{(\beta g)^{1/5}t^{2/5}}.
\label{eq:sft}
\end{equation}
Here, $\varepsilon_{\theta}(t)\equiv\langle\kappa[\partial_i\theta(x,z,t)]^2\rangle_V$ is the thermal dissipation rate and $\langle\cdots\rangle_V$ means a volume average inside the mixing zone. Extending relation (\ref{eq:sfu}) down to the Kolmogorov dissipation scale $\eta$, together with the relation $\delta v_{\eta}\eta\sim\nu$, one obtains
\begin{equation}
\eta(t)\sim(\frac{\delta v_{\eta}}{v_{rms}(t)})^{5/3}L(t)\sim\frac{\nu^{5/8}L(t)^{3/8}}{v_{rms}(t)}\sim\frac{\nu^{5/8}t^{1/8}}{(\beta g\Theta_0)^{1/4}}
\label{eq:eta}
\end{equation}
and then
\begin{equation}
\varepsilon_u(t)\sim\frac{\nu^3}{\eta(t)^4}\sim\frac{\beta g\Theta_0\nu^{1/2}}{t^{1/2}},
\label{eq:eu}
\end{equation}
where $\varepsilon_u(t)\equiv\langle\nu[\partial_iu_j(x,z,t)]^2\rangle_V$ is the kinetic-energy dissipation rate.

Relations (\ref{eq:force_balance})-(\ref{eq:eu}) are the main theoretical predictions of the Chertkov model for 2D RT turbulence. The spatial and temporal scaling, Eqs. (\ref{eq:sfu}) and (\ref{eq:sft}), have been numerically verified first by Celani \emph{et al.} \cite{celani2006prl} and then by a scale-by-scale study of Biferale \emph{et al.} \cite{biferale2010pof}. However, to the best of our knowledge, there are few studies concerning the other predictions, especially for the basic assumption of the force balance relation (\ref{eq:force_balance}). We remark that the quantitative test of Eq. (\ref{eq:force_balance}) is also required for turbulent RB convection and is considered to be a direct validation of the BO59 scenario \cite{lx2010arfm}. One of the objectives of the present paper is to validate these scaling predictions, i.e. relations (\ref{eq:force_balance}), (\ref{eq:et}), (\ref{eq:eta}), and (\ref{eq:eu}), in 2D miscible RT turbulence, on the basis of high-resolution direct numerical simulation. We further extend the dimensional predictions (\ref{eq:sfu}) and (\ref{eq:sft}) of the phenomenological theory to higher orders to include intermittency effects, which are beyond the mean-field theory.

\section{Numerical method}
In two dimensions, the vorticity-stream function formulation of Eq. (\ref{eq:ns_v}) is computationally advantageous for it eliminates the pressure variable and automatically enforces incompressibility. By introducing the vorticity $\omega=\nabla\times\mathbf{u}$ and the stream function $\psi$, Eq. (\ref{eq:ns_v}) is equivalent to
\begin{equation}
\frac{\partial \omega}{\partial t} + (\mathbf{u}\cdot\nabla)\omega=\nu\nabla^2\omega+\beta g\frac{\partial\theta}{\partial x},
\label{eq:ns_omega}
\end{equation}
\begin{equation}
\nabla^2\psi=\omega,
\label{eq:psi}
\end{equation}
\begin{equation}
u = -\frac{\partial\psi}{\partial z}, \mbox{\ \ }w=\frac{\partial\psi}{\partial x}.
\label{eq:uw}
\end{equation}

\begin{figure}[t]
  \centering
 \includegraphics[width=0.95\columnwidth]{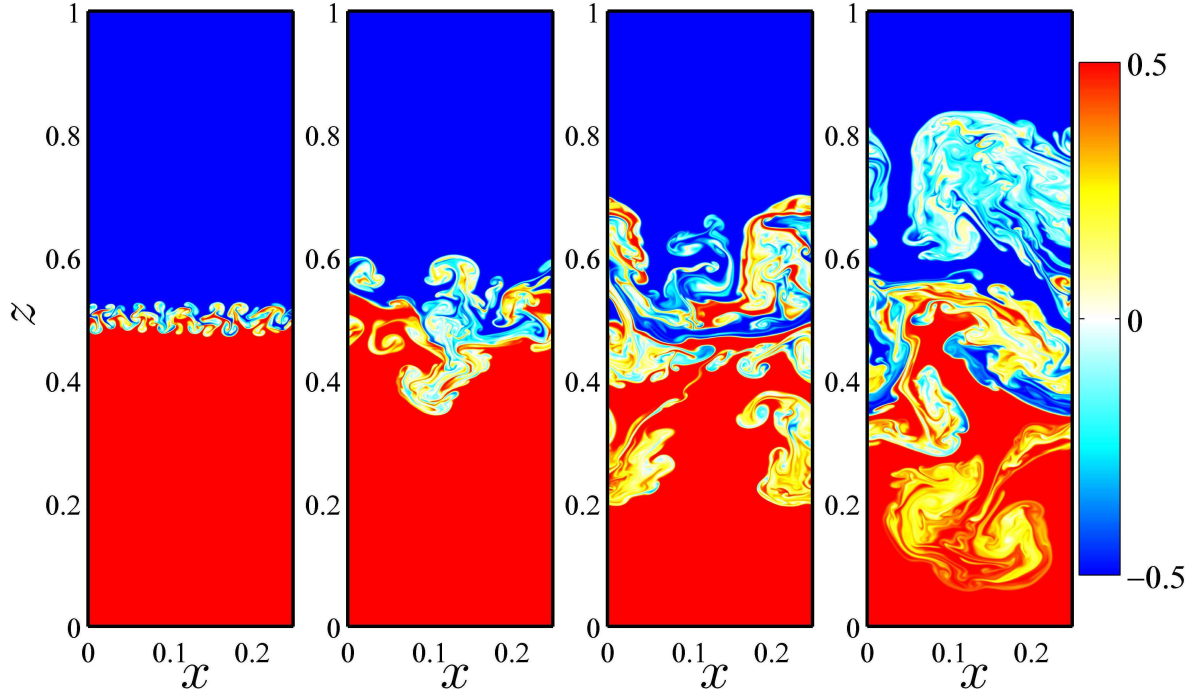}
  \caption{(Color online) Snapshots of the temperature fields for the RT evolution at $t/\tau=1$, 2, 3, and 4 (from left to right). Blue (red) regions correspond to cold (hot) fluid.}
  \label{fig:fig1}
\end{figure}

The numerical code adopted in this paper is based on a compact fourth-order finite difference scheme of the Oberbeck-Boussinesq equations (\ref{eq:ns_t}) and (\ref{eq:ns_omega})-(\ref{eq:uw}) on a 2D domain of width $L_x$ and height $L_z$ with uniform grid spacing $\Delta_g$. The scheme was proposed by Liu \emph{et al.} \cite{liu2003jsc}, who have examined the accuracy, stability, and efficiency of the scheme in great detail. This scheme has already been applied to the numerical studies of turbulent RB convection \cite{doering2009prl} and it was shown that the quantities obtained using this compact scheme agree well with those obtained using other numerical schemes, such as the Fourier-Chebyshev spectral collocation method \cite{doering2009prl}. Here we briefly describe the scheme. For Eqs. (\ref{eq:ns_omega})-(\ref{eq:uw}), an essentially compact fourth-order (EC4) scheme, first proposed by E $\&$ Liu \cite{liu1996jcp} for the 2D Navier-Stokes equations, is employed to solve the momentum equations with the gravity term treated explicitly. E $\&$ Liu \cite{liu1996jcp} have shown that the EC4 scheme has very nice features with regard to the treatment of boundary conditions. Such a scheme is also very efficient, because at each Runge-Kutta stage only two Poisson-like equations have to be solved by taking the standard fast fourier transform based Poisson solvers. The heat transfer equation (\ref{eq:ns_t}) is treated as a standard convection-diffusion equation and is discretized using fourth-order long-stencil difference operators. Third-order Runge-Kutta method is employed to integrate the equations in time. The time step is chosen to fulfill the Courant-Friedrichs-Lewy (CFL) conditions, i.e., the CFL number is 0.3 or less for all computations presented in this paper.

%\begin{table}
%\begin{center}
%\caption{Parameters of simulations: the number of grid points $N_x\times N_z$, Atwood number $A$, gravity $g$, viscosity $\nu$, thermal diffusivity $\kappa$, and the number of independent RT realizations $N_{ir}$.}
%\label{tab:tab1}
%\begin{tabular}{cccccccccccccccc}
%\hline
%\hline
%Label && $Ag$ && $N_x\times N_z$ && $L_x$ && $L_z$ && $\Theta_0$ && $\nu=\kappa$ && $N_{ir}$ \\
% &&&&&&&&&&&& $(\times10^{-6})$ && \\[4pt]
%\hline
%run A && 0.25 && $2048\times8193$ && 0.25 && 1 && 1 && 1.58 && 100\\
%run B && 0.1  && $2048\times4097$ && 0.5  && 1 && 1  && 3.16 && 60\\
%\hline
%\hline
%\end{tabular}
%\end{center}
%\end{table}

%Kinematic viscosity is chosen to be sufficiently large to ensure the resolution of small scales, i.e. $\eta>2\Delta_g$ for all time (see figure \ref{fig:fig10}). Two different sets of runs have been performed and Table \ref{tab:tab1} summarizes their parameters.

In the present study, the number of grid points is set to $2048\times8193$. Periodic boundary conditions are applied to the horizontal direction, while for the top and bottom plates, no-penetration and no-slip velocity boundary conditions, $\psi|_{z=-L_z/2,L_z/2}=0$ and $\partial\psi/\partial t|_{z=-L_z/2,L_z/2}=0$, and adiabatic (no flux) temperature boundary conditions are adopted. In all the runs, $Ag=0.25$, $L_z=1$, $\Theta_0=1$, $\nu=\kappa=1.58\times10^{-6}$, and thus the corresponding Prandtl number $Pr=\nu/\kappa=1$.

\begin{figure}[t]
  \centering
 \includegraphics[width=0.497\columnwidth]{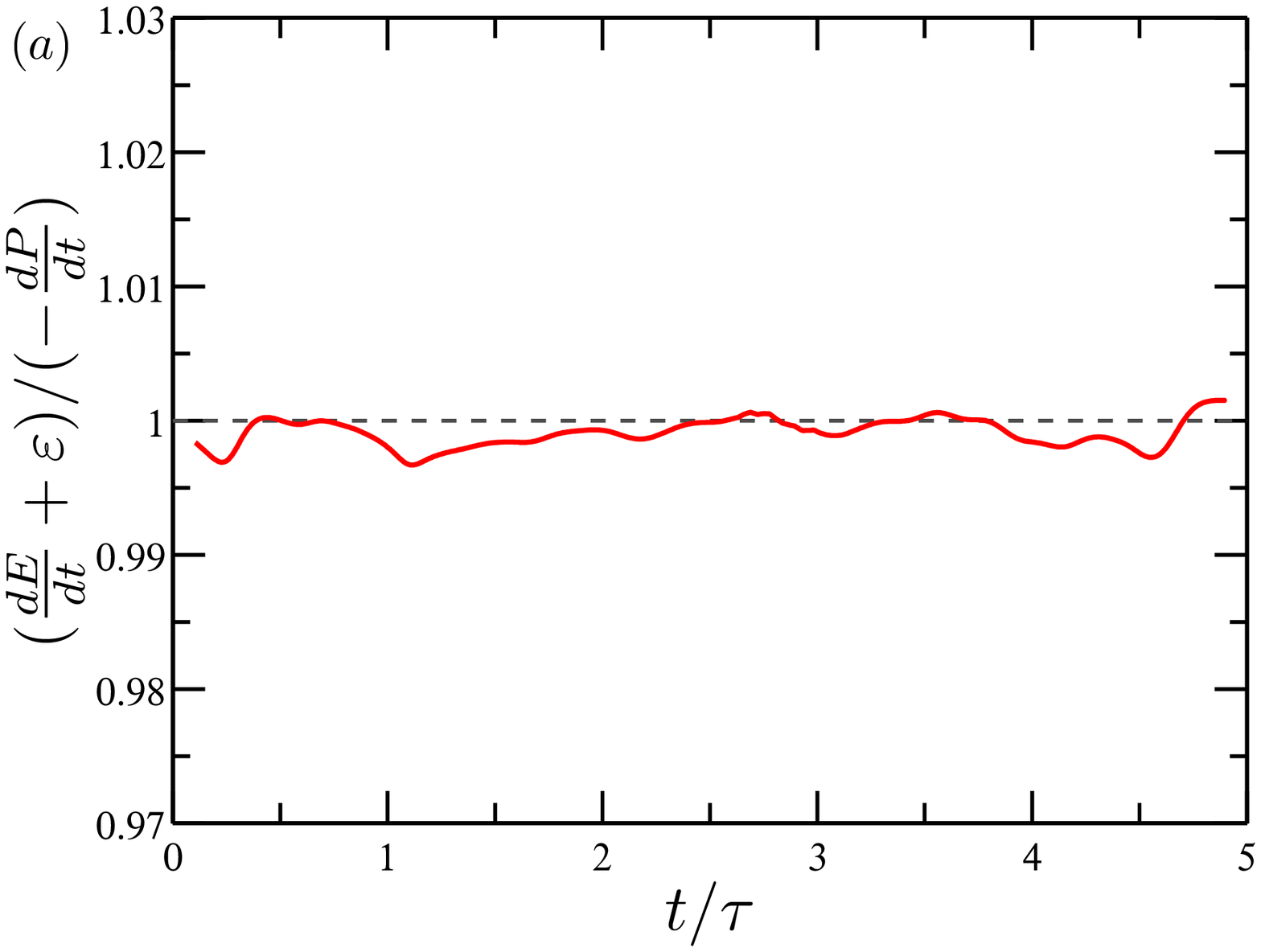}
 \includegraphics[width=0.492\columnwidth]{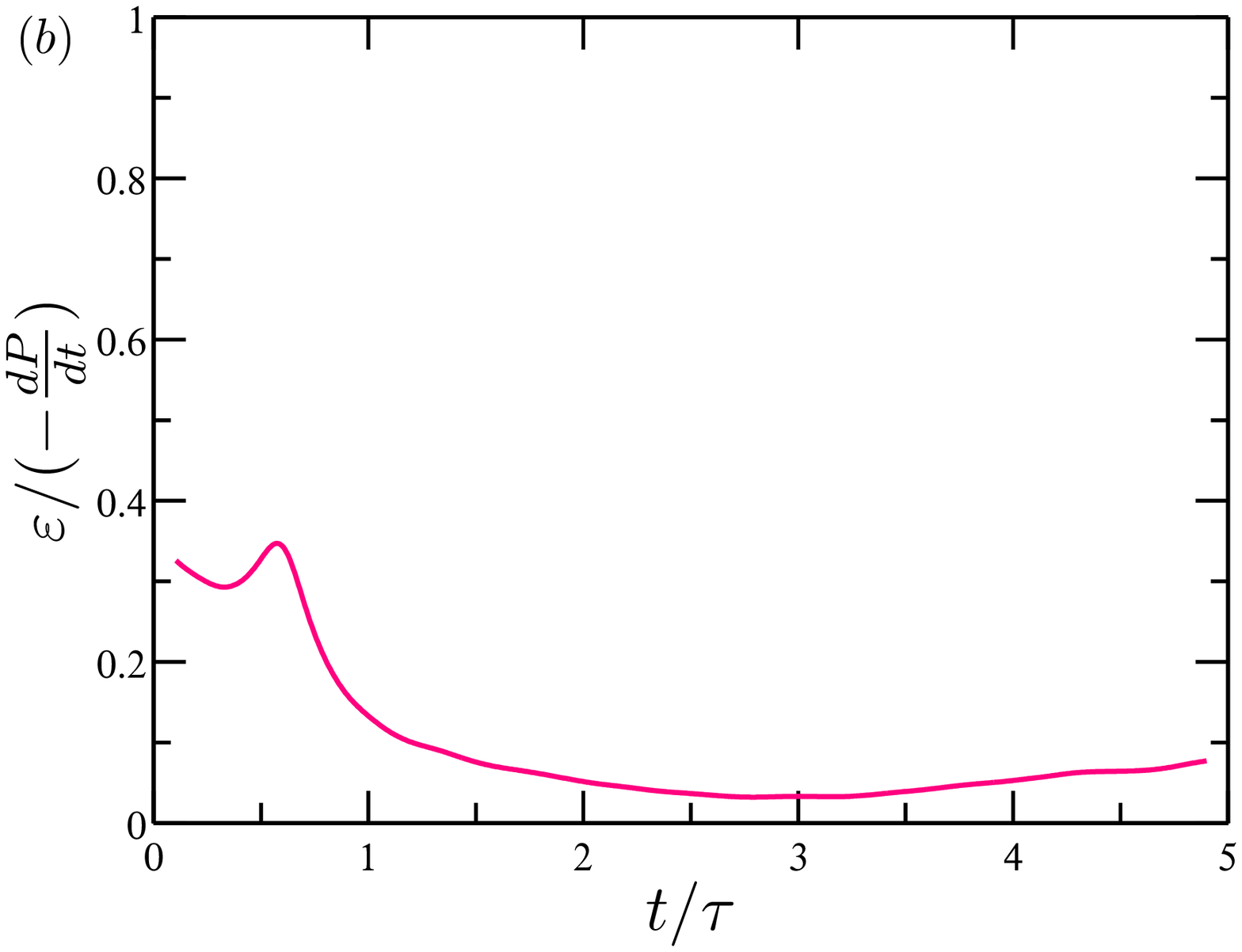}
  \caption{(Color online) (\emph{a}) Validation of the energy balance relation (\ref{eq:energy_balance}) for the numerical scheme. Here, $\mathrm{d}E/\mathrm{d}t$ is the total kinetic-energy growth rate, $\varepsilon$ is the total kinetic-energy dissipation rate, and $-\mathrm{d}P/\mathrm{d}t$ is the total potential-energy loss rate. (\emph{b}) The ratio of $\varepsilon$ to $-\mathrm{d}P/\mathrm{d}t$ as a function of time.}
  \label{fig:fig2}
\end{figure}

RT instability is seeded by adding a perturbed interface to the initial temperature interface $\theta=0$ at $z=0$. The perturbed interface is constructed from a superposition of cosine waves of wavenumbers $30\leq k\leq60$, equal amplitude, and random phases \cite{clark2003pof}. Figure \ref{fig:fig1} displays examples of the temperature fields at four distinct times during the RT evolution $t/\tau=1$, 2, 3, and 4, where $\tau=\sqrt{L_z/Ag}$ is the characteristic time of the RT system. One sees clearly that in the turbulence regime the flow is dominated by large-scale structures (plumes or spikes). Because the properties of these large-scale structures would show wide variations among individual simulations, a large number of statistically independent realizations must be calculated to assess the repeatability of the statistical quantities. In the present study, a total of 100 independent realizations of 2D RT evolution have been performed. In the remainder of this paper, all statistical quantities are obtained by first calculating for each individual simulation and then averaging over all these realizations. Simulations with the same parameters but a less resolution $1024\times4097$ are also performed. Comparison between the simulations of two different sizes suggests the robustness of the results in the present paper.

To validate the accuracy of the numerical code, we have checked the instantaneous kinetic energy budget relation:
\begin{equation}
-\frac{\mathrm{d}P(t)}{\mathrm{d}t}=\frac{\mathrm{d}E(t)}{\mathrm{d}t}+\varepsilon(t),
\label{eq:energy_balance}
\end{equation}
where $P(t)=-\beta g\int\int z\theta(x,z,t)\mathrm{d}x\mathrm{d}z$ is the total potential energy, $E(t)=\int\int(1/2)[u(x,z,t)^2+w(x,z,t)^2]\mathrm{d}x\mathrm{d}z$ is the total kinetic energy, and $\varepsilon(t)=\int\int\nu[\partial_iu_j(x,z,t)]^2\mathrm{d}x\mathrm{d}z$ is the total kinetic-energy dissipation rate. Figure \ref{fig:fig2}(\emph{a}) shows the ratio of the right-hand side to the left-hand side of Eq. (\ref{eq:energy_balance}) as a function of the normalized time $t/\tau$. It is seen that the energy balance equation is well verified within only $0.5\%$ at all times. In Fig. \ref{fig:fig2}(\emph{b}), we further plot the ratio of $\varepsilon(t)$ to $-\mathrm{d}P(t)/\mathrm{d}t$. While in the linear instability stage $\varepsilon(t)$ accounts for about $30\%$ of $-\mathrm{d}P(t)/\mathrm{d}t$, in the turbulence regime ($t\gtrsim\tau$) the dissipation rate $\varepsilon(t)$ becomes neglectable compared to the total potential-energy loss rate $-\mathrm{d}P(t)/\mathrm{d}t$. This amounts to saying that in the turbulence regime almost all the energy injected into the flow contributes to the growth of the large-scale flow. This is in clear contrast with the 3D case \cite{boffetta2010pof}, where an equipartition of large-scale energy growth and small-scale energy dissipation is observed.

\begin{figure}[t]
  \centering
 \includegraphics[width=0.6\columnwidth]{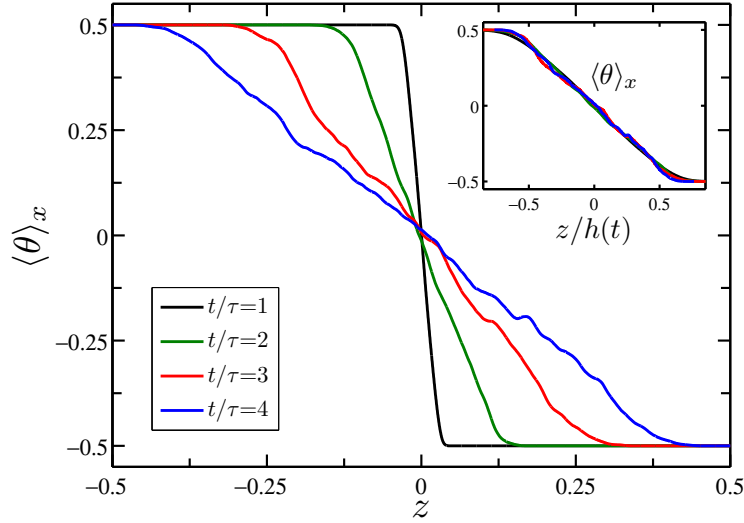}
  \caption{(Color online) Mean vertical temperature profiles $\langle\theta\rangle_x$ at times $t/\tau=1$, 2, 3, and 4. In the inset, the profiles are rescaled according to the instantaneous mixing zone width $h(t)$.}
  \label{fig:fig3}
\end{figure}

\section{Results and discussion}
\subsection{Global quantities}
Figure \ref{fig:fig3} displays the temporal evolution of the mean vertical temperature profiles $\langle\theta(x,z,t)\rangle_x$, where $\langle\cdots\rangle_x$ means a horizontal average. As observed in previous numerical studies \cite{celani2006prl, biferale2010pof}, the approximately linear behavior of the mean temperature profile can be seen within the mixing zone. Moreover, if the profiles are rescaled according to the instantaneous mixing zone width $h(t)$ as shown in the inset of Fig. \ref{fig:fig3}, all these profiles collapse well on top of each other. These suggest the self-similarity and homogeneity of the RT turbulent flow within the mixing zone in a statistical sense.

\begin{figure}[t]
  \centering
 \includegraphics[width=0.48\columnwidth]{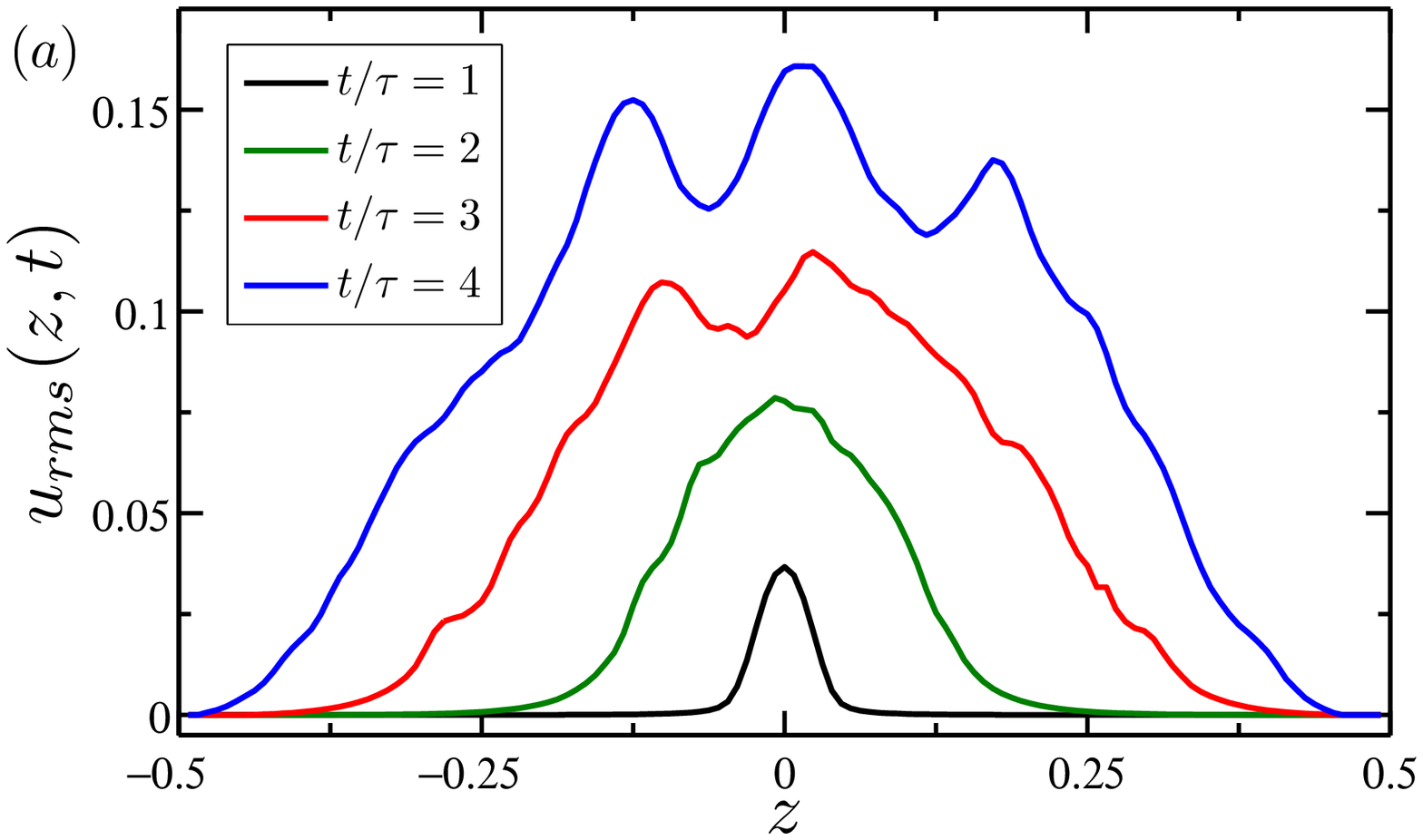}
 \includegraphics[width=0.48\columnwidth]{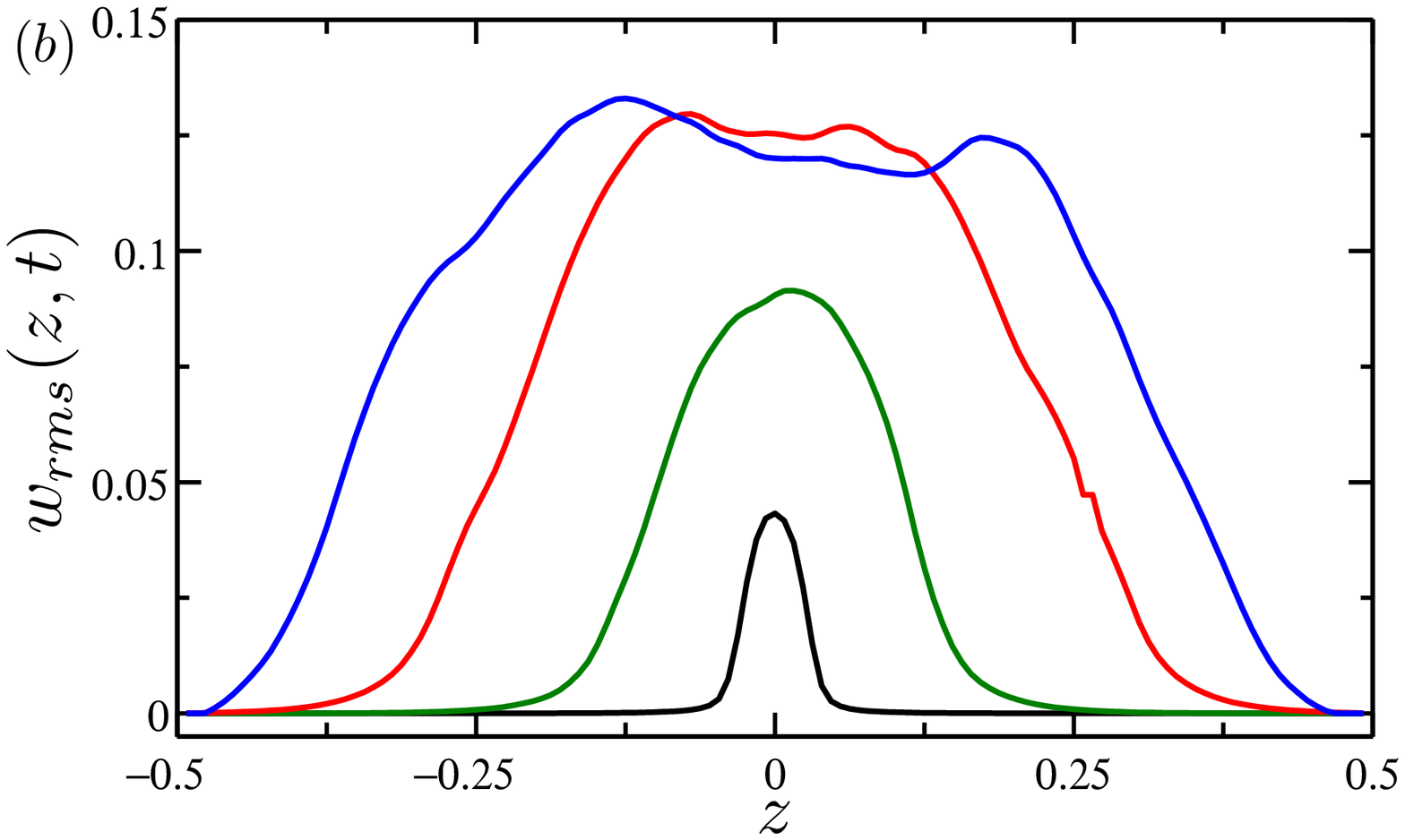}
 \includegraphics[width=0.48\columnwidth]{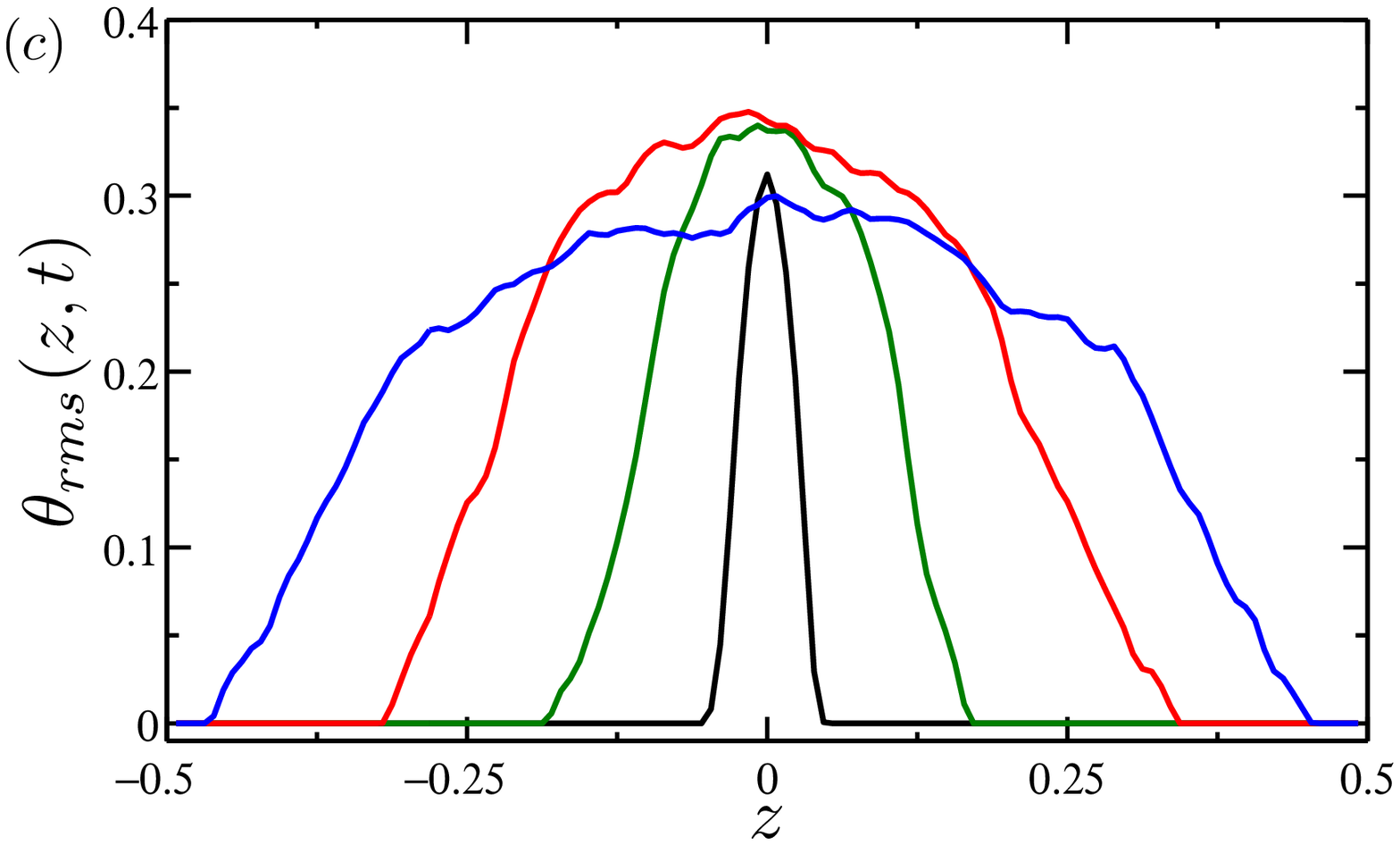}
 \includegraphics[width=0.482\columnwidth]{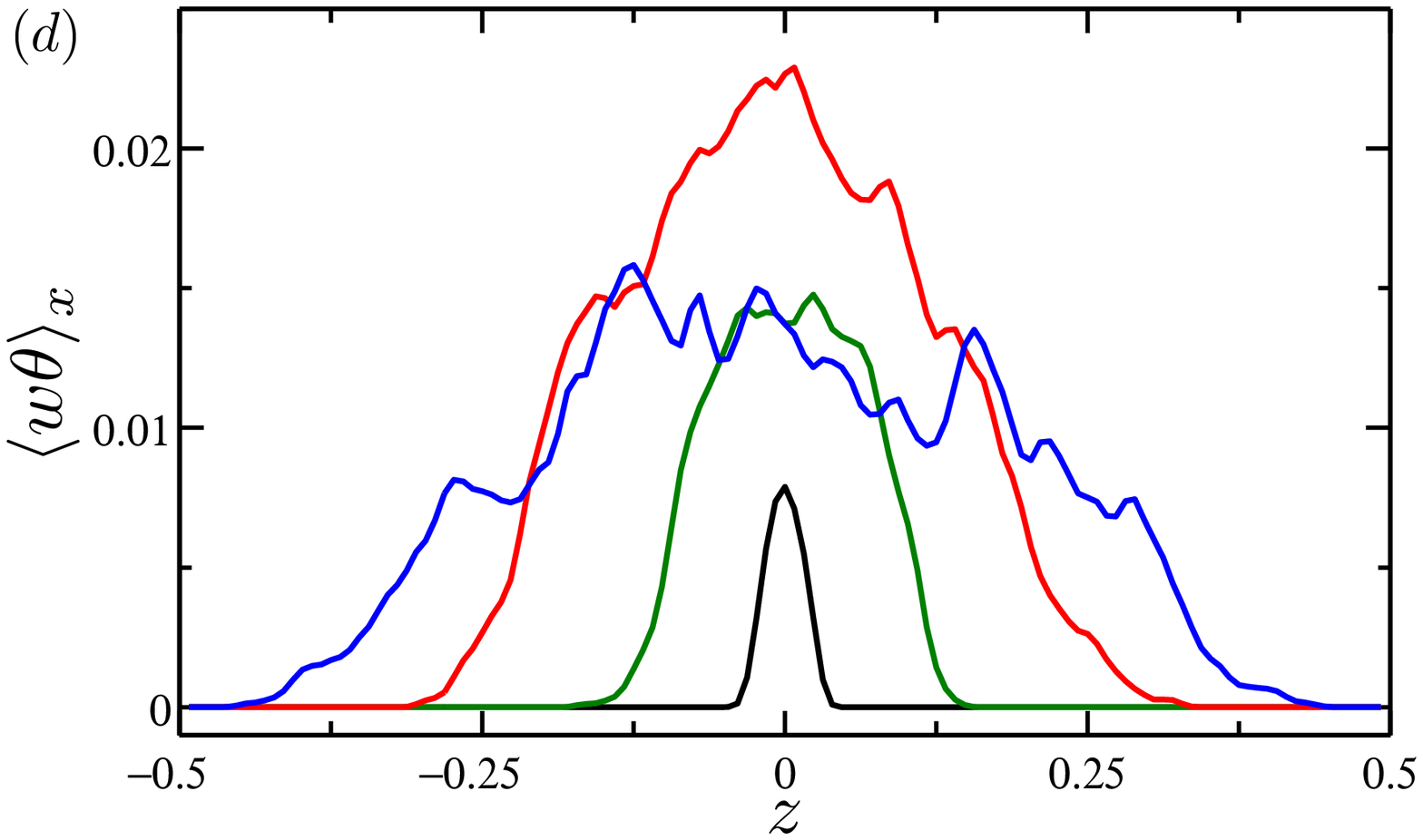}
  \caption{(Color online) Mean vertical profiles of the horizontal rms velocity $u_{rms}(z,t)$ (\emph{a}), the vertical rms velocity $w_{rms}(z,t)$ (\emph{b}), the rms temperature $\theta_{rms}(z,t)$ (\emph{c}), and the heat flux $\langle w\theta\rangle_x$ (\emph{d}) obtained at times $t/\tau=1$, 2, 3, and 4.}
  \label{fig:fig4}
\end{figure}

Figure \ref{fig:fig4} displays the temporal evolution of the profiles of the horizontal and vertical root-mean-square (rms) velocities $u_{rms}(z,t)$ and $w_{rms}(z,t)$, the rms temperature $\theta_{rms}(z,t)$, and the heat flux $\langle w\theta\rangle_x$, where $i_{rms}=\sqrt{\langle(i-\langle i\rangle_j)^2\rangle_j}$ is the rms value of $i$ with $i=u$, $w$, or $\theta$ and with $j=x$ for a horizontal average or $j=V$ for a volume average inside the mixing zone. For $t/\tau\lesssim3$ all these profiles show a similar shape, not far from a parabola, within the mixing zone. But, the time behaviors of the amplitudes of these four quantities are different. While the amplitudes of $u_{rms}(z,t)$, $w_{rms}(z,t)$, and $\langle w\theta\rangle_x$ increase almost linearly with time, the amplitude of $\theta_{rms}(z,t)$ keeps nearly constant as expected. For later times ($t/\tau$=4), the amplitudes of these quantities decrease with time except that of $u_{rms}(z,t)$ which still increases in time.

\begin{figure}[t]
  \centering
 \includegraphics[width=0.6\columnwidth]{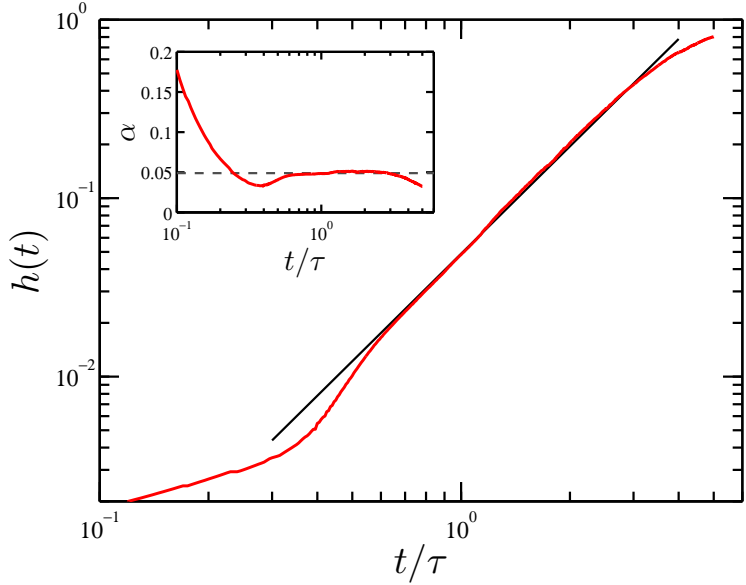}
  \caption{(Color online) Temporal evolution of the mixing layer width $h(t)$ computed with a threshold $s=0.8$. The straight line indicates the quadratic law $h(t)=\alpha Agt^2$ with $\alpha=0.049$ obtained from the compensated plot of $h(t)/(Agt^2)$ shown in the inset.}
  \label{fig:fig5}
\end{figure}

\begin{figure}[t]
  \centering
 \includegraphics[width=0.6\columnwidth]{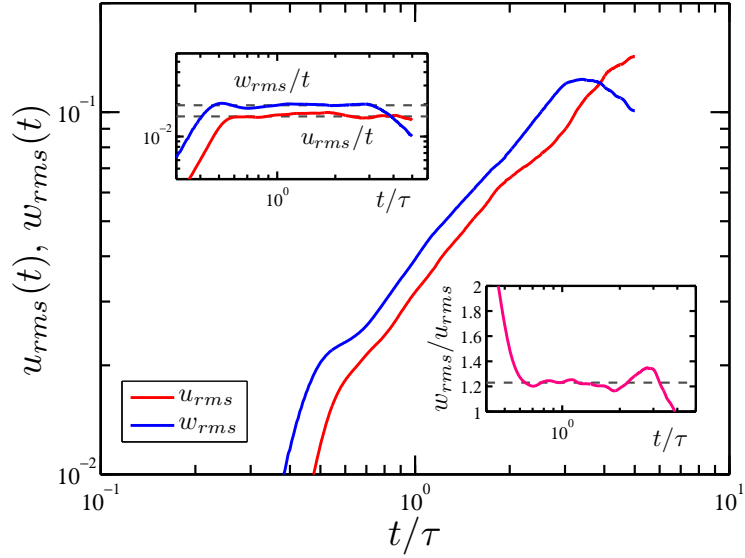}
  \caption{(Color online) Temporal evolution of the horizontal rms velocity $u_{rms}(t)$ (red line) and the vertical rms velocity $w_{rms}(t)$ (blue line). Upper inset: temporal evolution of $w_{rms}$ and $u_{rms}$ compensated with the linear scaling $t$. Lower inset: ratio of $w_{rms}$ to $u_{rms}$.}
  \label{fig:fig6}
\end{figure}

With the measured temperature profiles at different times during the RT evolution, one can define the width of the mixing zone, $h(t)$. There are mainly two sets of definitions of the mixing zone width \cite{andrews1990pof, youngs1999jfm, clark2003pof, cc2006np}: one is based on threshold values and the other on integral quantities. In the turbulence regime unrestricted by the boundaries of the computational domain, two sets of definitions of $h(t)$ are actually within an $O(1)$ systematic factor of each other. For simplicity, we adopt in the work the common definitions on the basis of a threshold value $s$, i.e. $\langle\theta\rangle_x(z=\pm h/2)=\mp s\Theta_0/2$ with $s=0.8$. The growth of the obtained mixing zone width $h(t)$ is plotted as a function of $t/\tau$ in Fig. \ref{fig:fig5}. The solid line in the figure represents the quadratic growth relation (\ref{eq:ht}) for $h(t)$ introduced in Sec. II. One sees that in the self-similarity regime $0.6\lesssim t/\tau\lesssim3$ the solid line describes well the temporal behavior of $h(t)$, indicating the $t^2$ scaling law for $h(t)$. This can be seen more clearly in the inset of Fig. \ref{fig:fig5}, where the compensated graph $h(t)/Agt^2$ becomes quite flat in the same range. From the evolution of $h(t)$, one is able to calculate the mixing zone growth rate $\alpha$. The value of $\alpha$ depends on the definition of the mixing zone width and has been studied extensively (see Ref. 10 for the review of the $\alpha$-studies). Here, the flat region in the inset of Fig. \ref{fig:fig5} yields 0.049 for the value of $\alpha$. We note that this value is larger than $\alpha=0.036$ found in both 3D \cite{boffetta2010pof} and quasi-2D \cite{boffetta2012jfm} cases using the same definition of $h(t)$ as our case, implying that the large-scale structures of 2D RT turbulence grow much faster due to the lack of the third dimensionality. For larger times ($t/\tau\gtrsim3$), the growth of $h(t)$ deviates from the square law. This deviation was also observed by Biferale \emph{et al.} \cite{biferale2010pof} using a thermal lattice Boltzmann method. We notice that $h(t)\gtrsim1.8L_x$ for $t/\tau\gtrsim3$ and thus this deviation is probably due to a transition from superdiffusive to subdiffusive evolution of the mixing zone \cite{boffetta2012pre}: when the vertical scale of the mixing zone $h(t)$ becomes much larger than its horizontal scale $L_x$, the lateral confinement prevents an efficient conversion of potential energy to vertical kinetic energy \cite{boffetta2012pre} (see also Fig. \ref{fig:fig6}).

Figure \ref{fig:fig6} shows the temporal evolution of the horizontal and vertical rms velocities $u_{rms}(t)$ and $w_{rms}(t)$, calculated inside the mixing zone. As discussed in Sec. II, the rms velocities are expected to grow linearly in time [see Eq. (\ref{eq:urms})]. Here, we indeed observe a linear growth for both $u_{rms}(t)$ and $w_{rms}(t)$ in the self-similarity regime $0.6\lesssim t/\tau\lesssim3$ (see the upper inset of Fig. \ref{fig:fig6} for a compensated plot). For later times ($t/\tau\gtrsim3$), the horizontal rms velocity $u_{rms}(t)$ continues to increase linearly with time, while the growth of the vertical one $w_{rms}(t)$ is prohibited. This again implies a confinement induced transition from accelerated to subdiffusive. As already discussed, when the aspect ratio of the mixing zone $L_x/h(t)$ becomes smaller than unity, the lateral confinement arrests the growth of the vertical kinetic energy, leading to the increase of the horizontal kinetic energy at the expenses of the potential energy. Another feature worthy of note is that in the self-similarity regime the vertical rms velocity $w_{rms}$ is about 1.2 times larger than the horizontal one $u_{rms}$ (see the lower inset of Fig. \ref{fig:fig6}), reflecting the anisotropy of the forcing due to gravity. We further notice that the ratio $w_{rms}/u_{rms}$ obtained here is smaller than the value 1.8 for the 3D case \cite{boffetta2010pof}, implying a relatively lower degree of large-scale anisotropy possessed by 2D RT turbulence.

\begin{figure}[t]
  \centering
 \includegraphics[width=0.6\columnwidth]{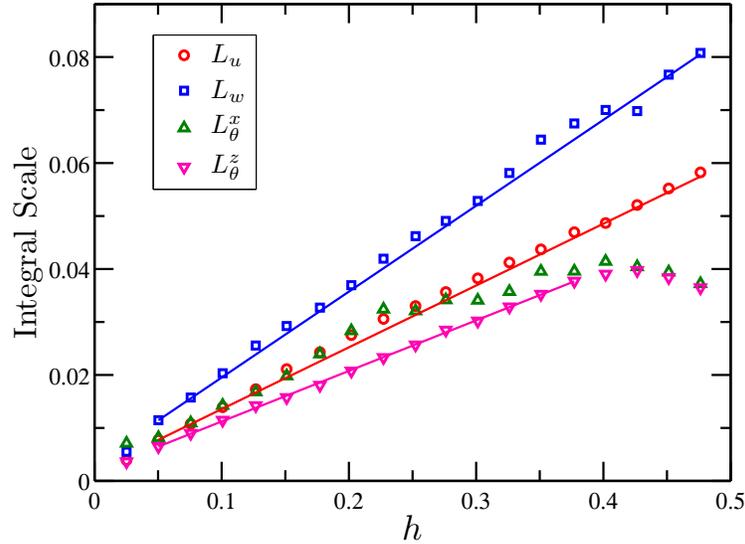}
  \caption{(Color online) Growth of the integral length scales. $L_u$ (circles) and $L_w$ (squares) correspond to scales computed using the horizontal and vertical components of the velocity along the horizontal and vertical directions, respectively. $L_{\theta}^x$ (up-triangles) and $L_{\theta}^z$ (down-triangles) correspond to scales computed using temperature along the horizontal and vertical directions, respectively. The solid lines are the best linear fits to the corresponding data.}
  \label{fig:fig7}
\end{figure}

\begin{figure}[t]
  \centering
 \includegraphics[width=0.495\columnwidth]{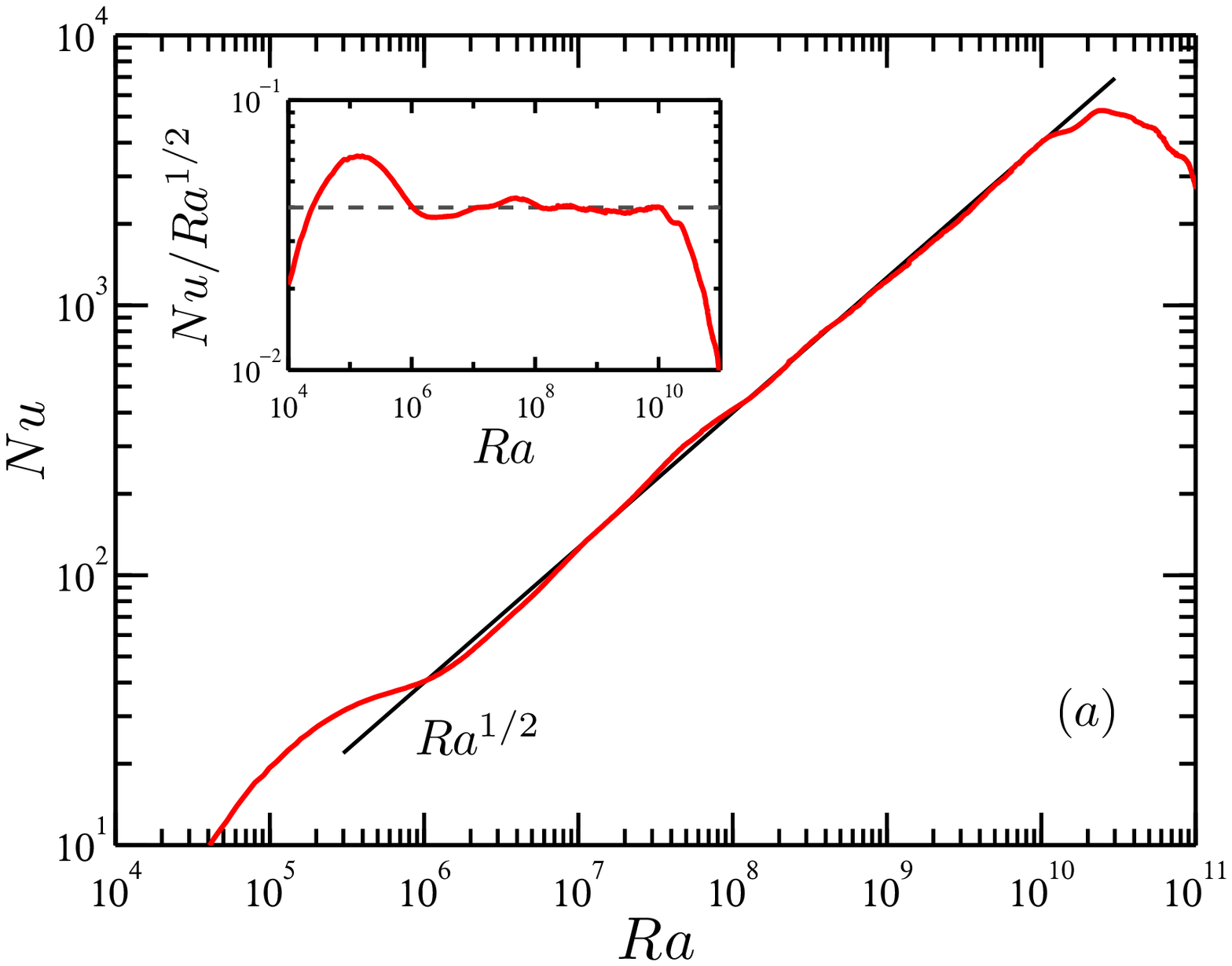}
 \includegraphics[width=0.495\columnwidth]{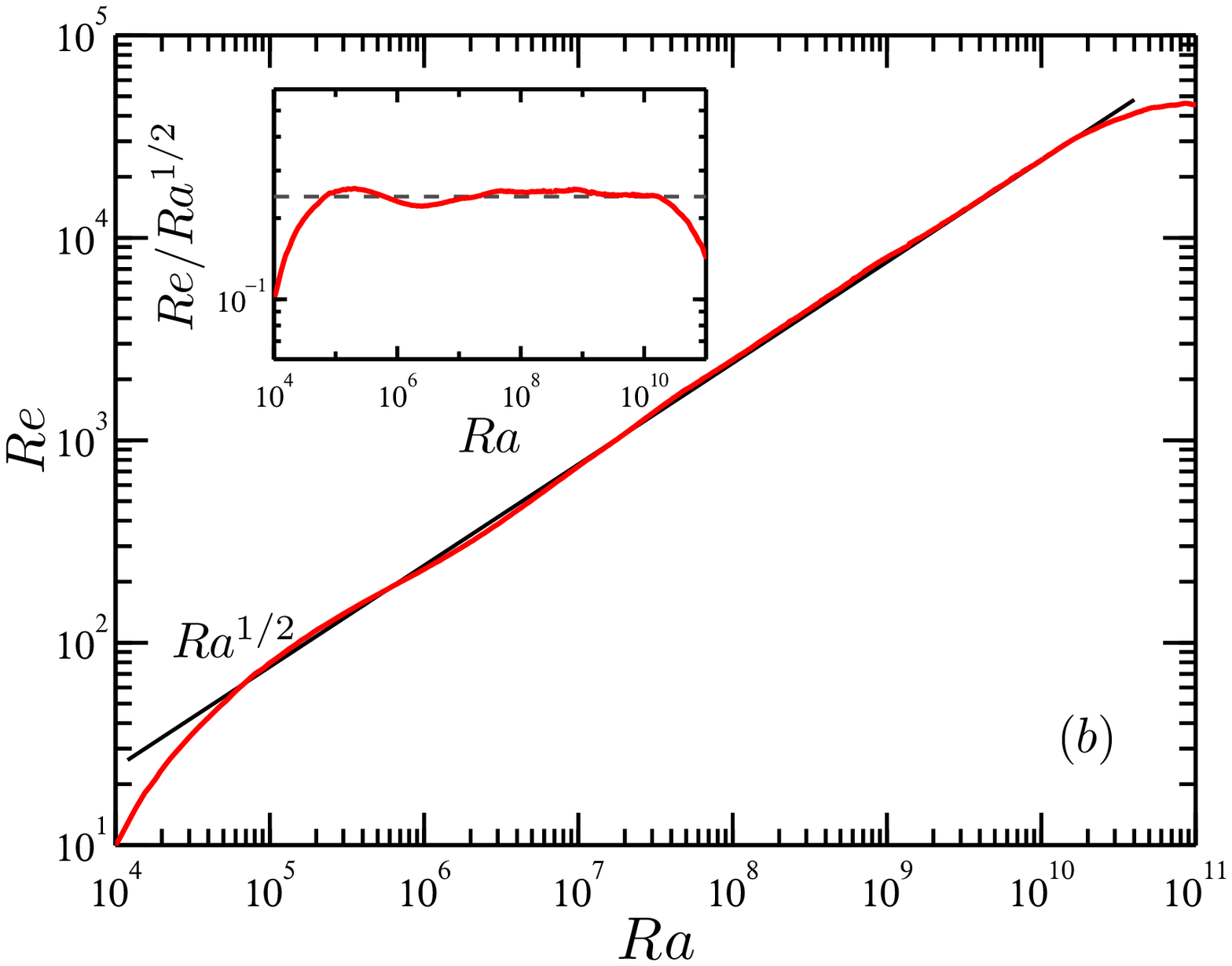}
  \caption{(Color online) The behaviors of the Nusselt number $Nu$ (\emph{a}) and the Reynolds number $Re$ (\emph{b}) versus the Rayleigh number $Ra$. Insets: $Nu$ and $Re$ compensated with the ultimate-state scaling $Ra^{1/2}$.}
  \label{fig:fig8}
\end{figure}

The characteristic size of large-scale turbulent eddies is usually expressed in terms of the integral length scale $L(t)$. As RT turbulence is a paradigmatic example of time-dependent turbulent system, the integral scale of the RT system is expected to grow in time linearly with the evolution of the mixing zone [see Eq. (\ref{eq:lt})]. For the 3D case, this was indeed observed by previous numerical studies \cite{chertkov2009pof, boffetta2010pof}. In this study, we adopt the idea of Vladimirova $\&$ Chertkov \cite{chertkov2009pof} to investigate the relation between $L(t)$ and $h(t)$ in 2D RT flow. We consider the integral scales $L_u$ and $L_w$ for the velocity field and $L_{\theta}^x$ and $L_{\theta}^z$ for the temperature field. Here, $L_u$ and $L_w$ are estimated as the half width of the horizontal and vertical velocity correlation functions $f(L_u)=\langle u(x,z)u(x+L_u,z)\rangle_V/u_{rms}^2=0.5$ and $f(L_w)=\langle w(x,z)w(x,z+L_w)\rangle_V/w_{rms}^2=0.5$, respectively, and $L_{\theta}^x$ and $L_{\theta}^z$ are estimated as the half width of the temperature correlation functions $f(L_{\theta}^x)=\langle \theta(x,z)\theta(x+L_{\theta}^x,z)\rangle_V/\theta_{rms}^2=0.5$ and $f(L_{\theta}^z)=\langle \theta(x,z)\theta(x,z+L_{\theta}^z)\rangle_V/\theta_{rms}^2=0.5$, respectively. As shown in Fig. \ref{fig:fig7}, we observe in the turbulent regime a linear relation between the integral scales and the mixing zone width. For velocity, a best linear fit yields $L_u/h\simeq0.12$ and $L_w/h\simeq0.16$ for the integral scales based on the horizontal and vertical components of the velocity, respectively. These values are much larger than those of $L_u/h\simeq0.024$ and $L_w/h\simeq0.059$ obtained in 3D case \cite{chertkov2009pof, boffetta2010pof}, suggesting that the 2D large-scale vortices in the mixing zone grow much faster than their 3D counterparts. In addition, the relative difference between $L_u/h$ and $L_w/h$ of the present case is smaller than that of the 3D case, suggesting again a relatively lower degree of anisotropy possessed by 2D RT flow. For temperature, in the horizontal direction $L_\theta^x$ first follows the growth of the $L_u$, reaches its maximum value around $h(t)\simeq0.4$, and then drops slightly at later times. In the vertical direction, $L_\theta^z$ shares the similar $h$-dependence with a smaller slope $L_\theta^z/h\simeq0.095$.

\begin{figure}[t]
  \centering
 \includegraphics[width=0.6\columnwidth]{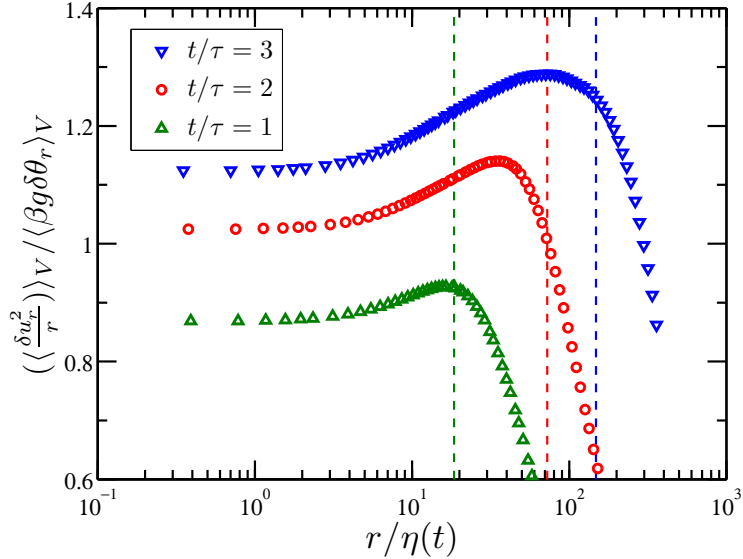}
  \caption{(Color online) Ratio of the inertial force $\langle\delta u_r^2/r\rangle_V$ to the buoyancy force $\langle\beta g\delta\theta_r\rangle_V$ as a function of the normalized scale $r/\eta(t)$ at times $t/\tau=1$, 2, and 3 during the RT evolution. The vertical dashed lines mark the corresponding integral length scales.}
  \label{fig:fig9}
\end{figure}

At the end of this section, we discuss the relations of the turbulent heat flux and the kinetic energy to the mean temperature gradient in their dimensionless form, i.e. the Nusselt number $Nu=1+\langle w\theta\rangle_Vh/\kappa\Theta_0$, the Reynolds number $Re=\sqrt{u_{rms}^2+w_{rms}^2}h/\nu$, and the Rayleigh number $Ra=\beta g\Theta_0h^3/\nu\kappa$, respectively. Figure \ref{fig:fig8} shows the measured Nusselt and Reynolds versus Rayleigh laws. A clear scaling can be seen for both $Nu(Ra)$ and $Re(Ra)$ for nearly four decades from $Ra\simeq10^6$ to $10^{10}$. The compensated plots in the insets give
\begin{equation}
Nu\sim Ra^{0.5} \mbox{\ \ and\ \ } Re\sim Ra^{0.5}.
\label{eq:ultimate}
\end{equation}
The relation (\ref{eq:ultimate}) represents the so-called ultimate state regime of RT turbulence \cite{boffetta2012pd}, which has been observed in both 2D \cite{celani2006prl, biferale2010pof, boffetta2012pd} and 3D \cite{boffetta2009pre, boffetta2010pof, boffetta2012pd} numerical simulations. Notice that the ultimate state scaling was first proposed for turbulent RB convection at very high Rayleigh number by Kraichnan, and then retrieved by Grossmann $\&$ Lohse in their $Ra$-$Pr$ phase diagram where the bulk turbulence dominates both the global  kinetic and thermal dissipation of the system \cite{agl2009rmp}. Although the existence of the ultimate state is still an open issue for the traditional RB convection with solid boundaries \cite{ahlers2012prl, urban2012prl}, it has been shown, both numerically \cite{lohse2003prl} and experimentally \cite{castaing2006prl, shang2008prl}, that the ultimate regime scaling can be realized when the solid boundaries are absent. In RT turbulence, the observation of the ultimate state is not surprising as boundaries play no role in the system and the bulk dynamics dominate the convective turbulence in the mixing zone. We remark that as shown in Fig. \ref{fig:fig8} the ultimate scaling regime $10^6\lesssim Ra\lesssim10^{10}$ corresponds to the self-similarity regime of the mixing zone evolution $0.6\lesssim t/\tau\lesssim3$ in which the accelerated law $h(t)\sim t^2$ is well established. For later times ($Ra>10^{10}$) the growth of $Nu$ is reduced and the $Re$-$Ra$ relation deviates from the scaling relation (\ref{eq:ultimate}). These deviations can be probably explained by the confinement induced transition \cite{boffetta2012pre} as discussed above.

\subsection{Small-scale properties}

\begin{figure}[t]
  \centering
 \includegraphics[width=0.6\columnwidth]{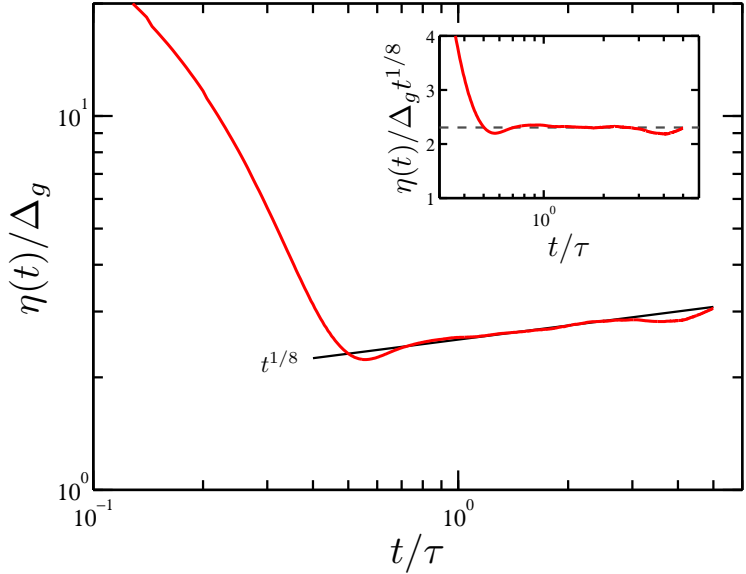}
  \caption{(Color online) Temporal evolution of the Kolmogorov scale $\eta(t)$ normalized by the computational grid spacing $\Delta_g$. The solid line is the temporal scaling prediction $t^{1/8}$ for reference. Inset: the compensated Kolmogorov scale $\eta(t)/(\Delta_g t^{1/8})$ as a function of time.}
  \label{fig:fig10}
\end{figure}

Let's now turn to the discussion of small-scale properties in 2D RT turbulence. Here, we mainly focus on the theoretical predictions of the Chertkov's 2D model \cite{chertkov2003prl}, i.e. Eqs. (\ref{eq:force_balance}), (\ref{eq:et}), (\ref{eq:eta}), and (\ref{eq:eu}). As already introduced in Sec. II, the basic assumption of the BO59 scenario is the force balance relation (\ref{eq:force_balance}) between the buoyancy force and the inertial force at all scales in the inertial subrange. In Fig. \ref{fig:fig9}, we plot the ratio of the inertial force $\langle\delta u_r^2/r\rangle_V$ to the buoyancy force $\langle\beta g\delta\theta_r\rangle_V$ as a function of the normalized scale $r/\eta(t)$ for three distinct times in the self-similarity regime. Here, $\delta\theta_r=|\theta(x+r,z)-\theta(x,z)|$ and $\delta u_r=|u(x+r,z)-u(x,z)|$ are temperature and horizontal velocity differences over a horizonal separation $r$, respectively. The vertical dashed lines in the figure mark the integral length scale $L_u$ of the horizontal velocity as shown in Fig. \ref{fig:fig7}. One sees that the ratio is close to unity for all scales below the integral one for all three sets of data, implying an approximate equipartition between the buoyancy force and the inertial force and thus validating the force balance relation (\ref{eq:force_balance}). Furthermore, it is seen that the ratio becomes larger at increasing time, indicating an increased magnitude of the inertial force with respect to the buoyancy force. The inertial force based on vertical velocity differences is also computed and compared with the corresponding buoyancy force. The similar results (not shown) are obtained.

\begin{figure}[t]
  \centering
 \includegraphics[width=0.495\columnwidth]{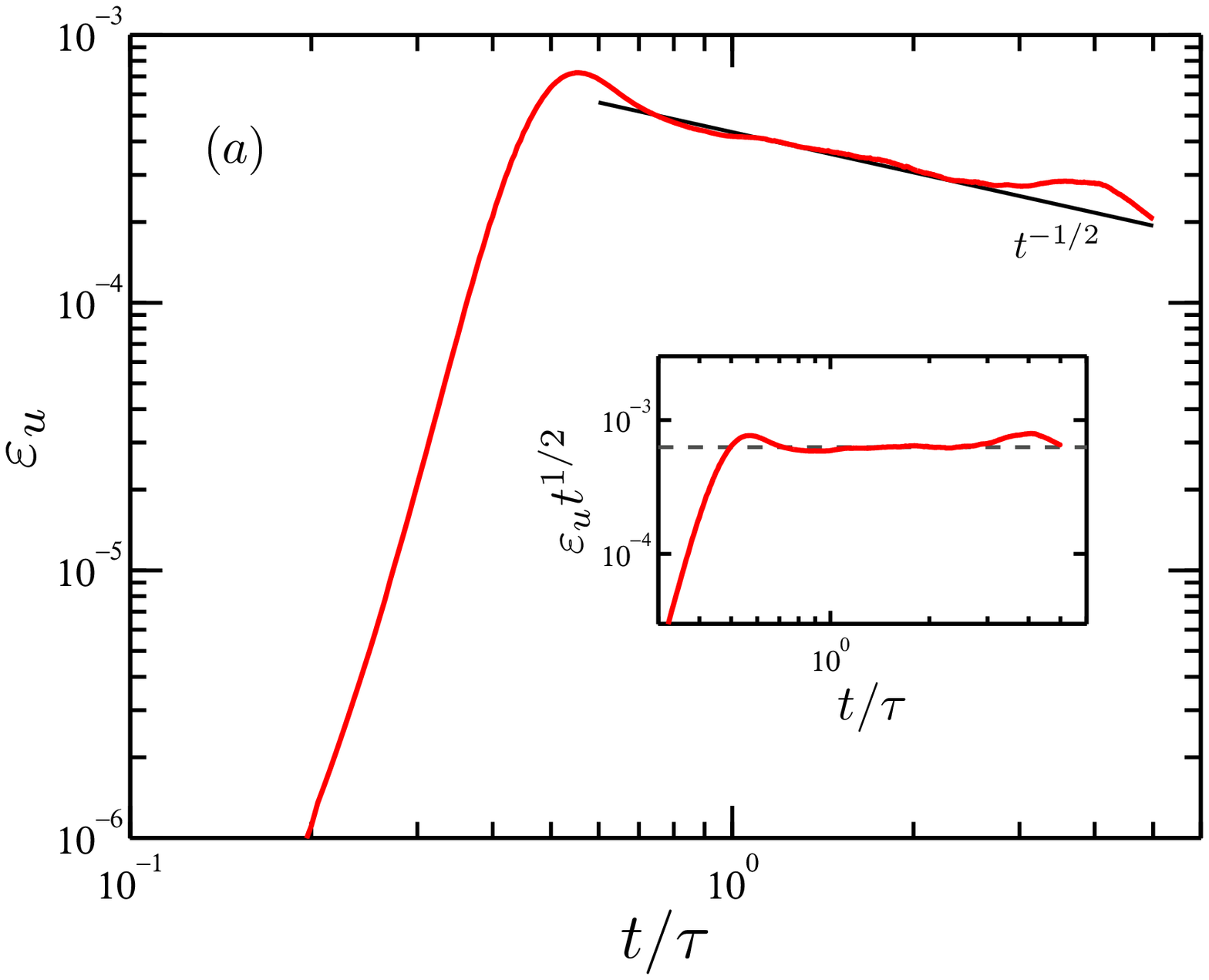}
 \includegraphics[width=0.495\columnwidth]{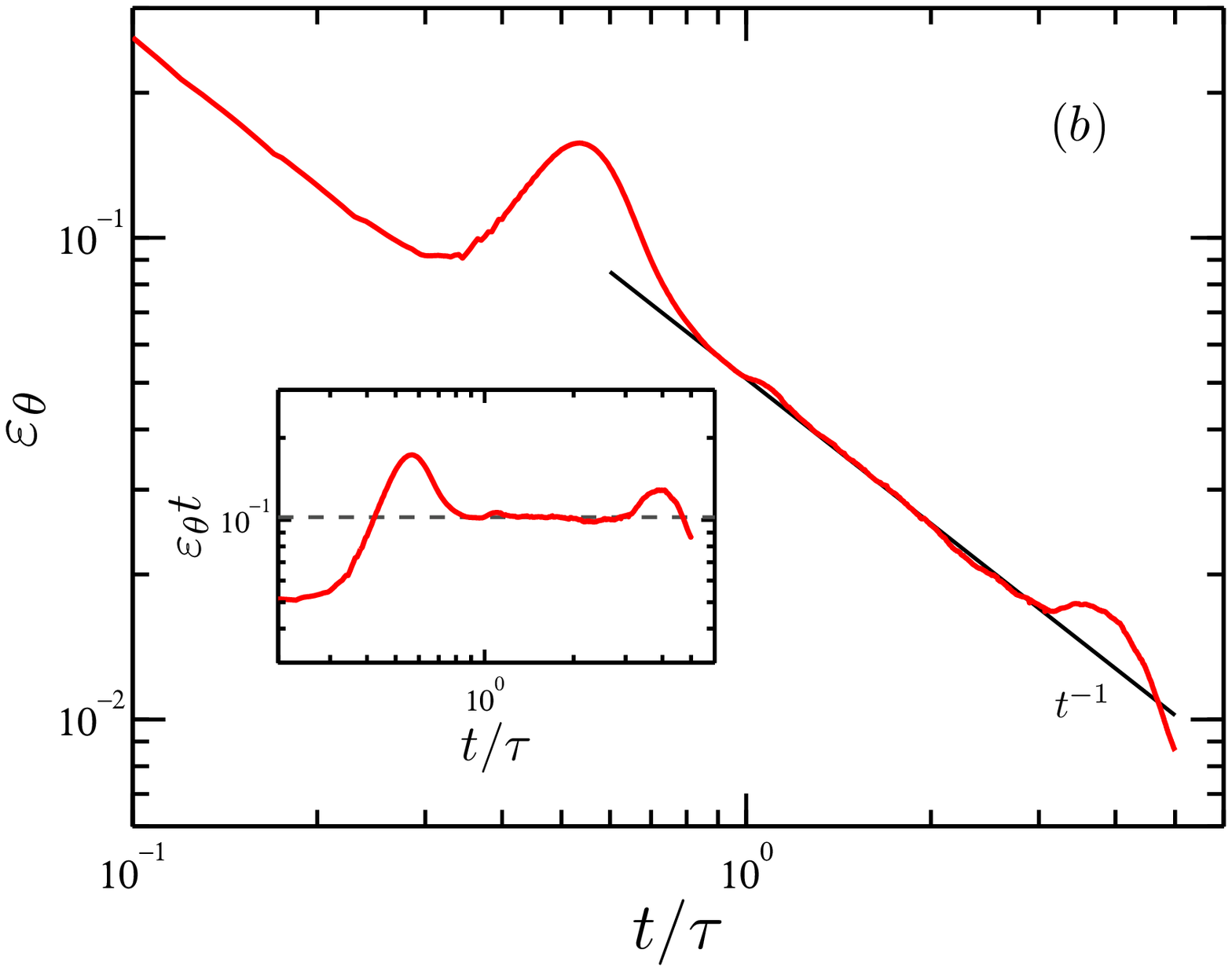}
  \caption{(Color online) Temporal evolution of the kinetic-energy dissipation rate $\varepsilon_u$ (\emph{a}) and the thermal dissipation rate $\varepsilon_{\theta}$ (\emph{b}). The solid lines are the temporal scaling predictions $t^{-1/2}$ and $t^{-1}$ for reference. Insets: the corresponding compensated $\varepsilon_ut^{1/2}$ and $\varepsilon_{\theta}t$.}
  \label{fig:fig11}
\end{figure}

Figure \ref{fig:fig10} shows the temporal evolution of the Kolmogorov dissipation scale $\eta(t)$, normalized by the grid spacing $\Delta_g$. It is seen that with increasing time $\eta(t)$ first drops quickly, reaches its minimum value at $t/\tau\simeq0.55$, and then rises slightly. The compensated plot in the set of Fig. \ref{fig:fig10} gives $\eta(t)\sim t^{1/8}$ in the self-similarity regime and thus validates Eq. (\ref{eq:eta}). This is at clear variance with the 3D case \cite{chertkov2003prl, cc2006np}, in which the Kolmogorov scale $\eta(t)$ decreases with time as $\eta(t)\sim t^{-1/4}$. During the simulations, we observe $\eta(t)>2\Delta_g$ for all times, guaranteeing the resolution of small scales.

The time behaviors of the kinetic-energy and thermal dissipation rates, $\varepsilon_u(t)$ and $\varepsilon_\theta(t)$, are plotted in Fig. \ref{fig:fig11} (\emph{a}) and (\emph{b}), respectively. In the self-similarity regime, both $\varepsilon_u(t)$ and $\varepsilon_\theta(t)$ decrease with time. The solid lines in the figure mark the temporal scaling predictions $\varepsilon_u(t)\sim t^{-1/2}$ and $\varepsilon_\theta(t)\sim t^{-1}$ for reference. One sees that both quantities follow the theoretical predictions well, which can be seen more clearly in their compensated forms (see the insets of Fig. \ref{fig:fig11}), thus implying the validation of Eqs. (\ref{eq:et}) and (\ref{eq:eu}). Note that the time-dependence of $\varepsilon_{\theta}$, $\varepsilon_{\theta}\sim t^{-1}$, is yielded from the scale-independent thermal balance of Eq. (\ref{eq:ns_t}) and thus is valid for both 2D and 3D cases no matter whether temperature is active or passive. While the relation $\varepsilon_u\sim t^{-1/2}$ holds only for the 2D case. It is in contrast with the 3D case \cite{boffetta2010pof}, where $\varepsilon_u(t)$ increases linearly with time $t$. It also differs slightly from the quasi-2D case \cite{boffetta2012jfm}, where the matching of the K41 and BO59 scalings gives $\varepsilon_u(t)\sim t^{-3/5}$.

%\begin{figure}[t]
%  \centering
% \includegraphics[width=0.495\columnwidth]{fig_pdfa.eps}
% \includegraphics[width=0.495\columnwidth]{fig_pdfb.eps}
%  \caption{(Color online) Tails of the PDFs of the kinetic-energy (\emph{a}) and thermal (\emph{b}) dissipation rates, $\varepsilon_u$ and $\varepsilon_{\theta}$, normalized by their respective rms values, $(\varepsilon_u)_{rms}$ and $(\varepsilon_{\theta})_{rms}$. The tails have been fitted by stretched exponentials (solid lines) as given by Eq. (\ref{eq:stretched}) with the fit exponents as indicated in the figure. For clarity, in (\emph{a}) and (\emph{b}) the $t/\tau=2$ and 3 data have been shifted upwards by 3.3 and 10, respectively, with respect to the $t/\tau=1$ data. The insets replot the same data in log-normal coordinates. The dashed lines indicate a log-normal distribution for reference.}
%  \label{fig:fig_pdf}
%\end{figure}
%
%\begin{equation}
%p(X^*)=\frac{C}{\sqrt{X^*}}\exp(-aX^{*\gamma}),
%\label{eq:stretched}
%\end{equation}

\begin{figure}[t]
  \centering
 \includegraphics[width=0.495\columnwidth]{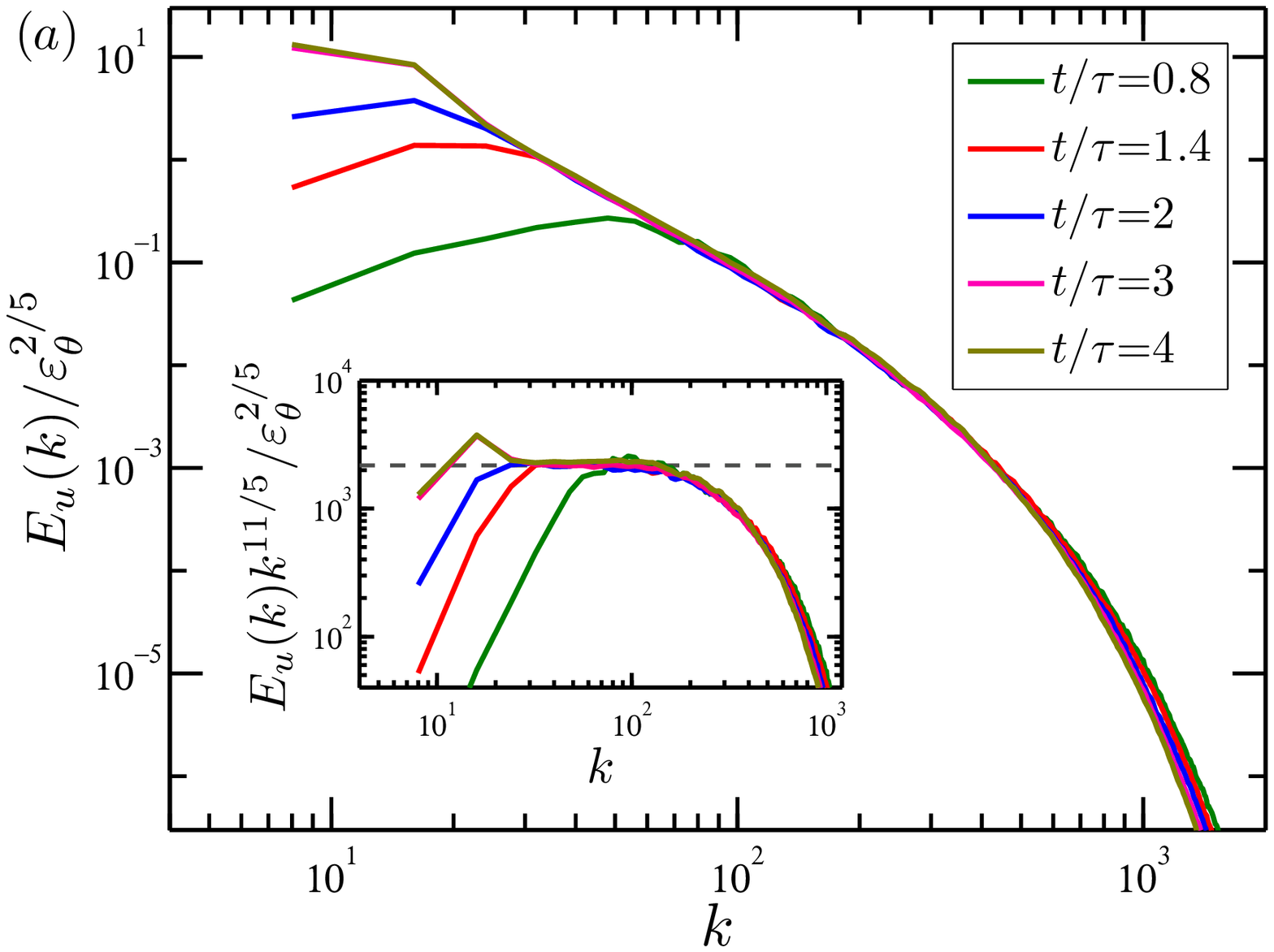}
 \includegraphics[width=0.495\columnwidth]{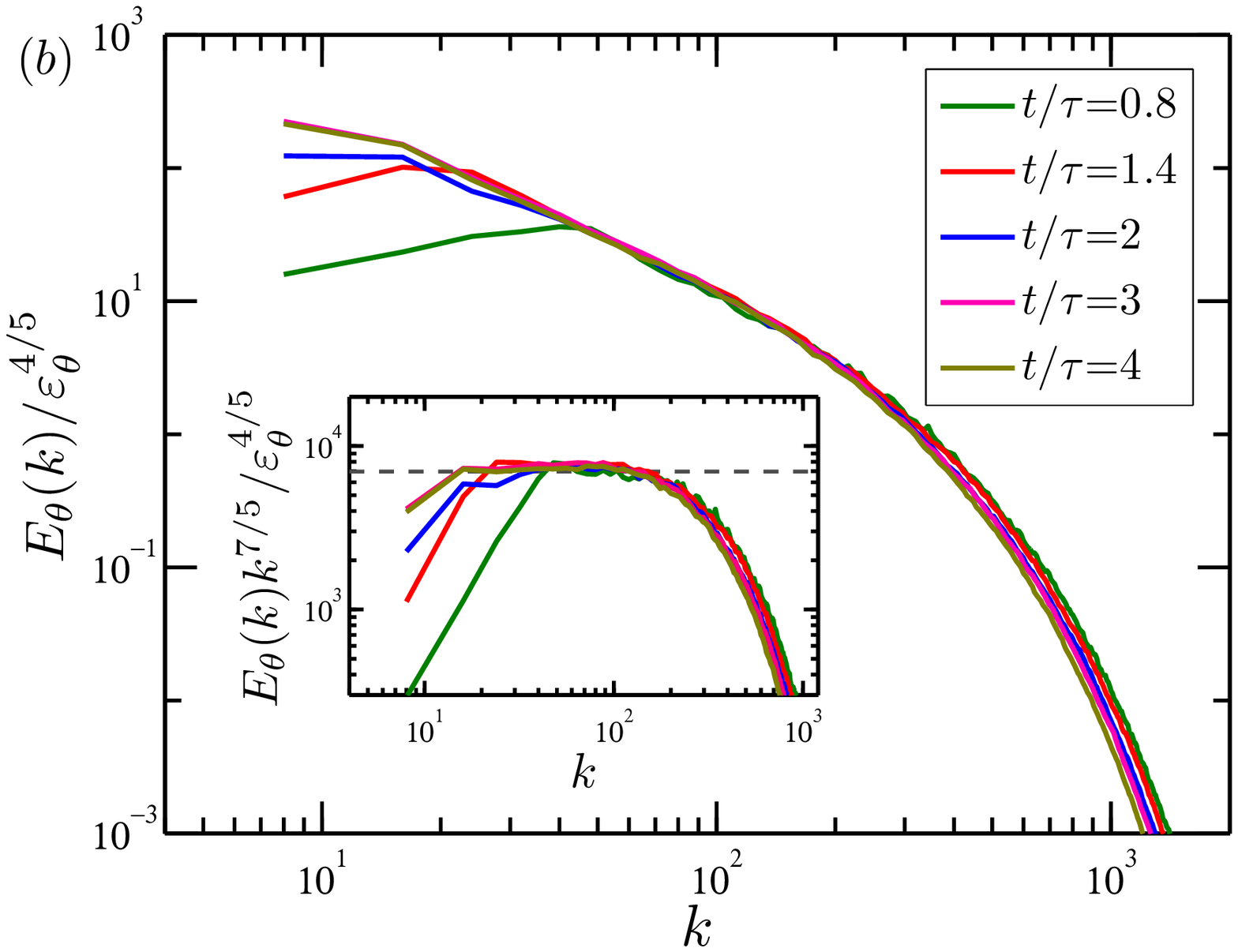}
  \caption{(Color online) (\emph{a}) Velocity power spectra $E_u(k)$ compensated with $\varepsilon_{\theta}^{2/5}$ at times $t/\tau=0.8$, 1.4, 2, 3, and 4. Inset: $E_u(k)$ compensated with the BO59 velocity scaling prediction $\varepsilon_{\theta}^{2/5}k^{-11/5}$. (\emph{b}) Temperature power spectra $E_{\theta}(k)$ compensated with $\varepsilon_{\theta}^{4/5}$ at the same times. Inset: $E_{\theta}(k)$ compensated with the BO59 temperature scaling prediction $\varepsilon_{\theta}^{4/5}k^{-7/5}$.}
  \label{fig:fig12}
\end{figure}

We next look at the BO59 scaling (\ref{eq:sfu}) and (\ref{eq:sft}) in Fourier space, where the corresponding spectra for the kinetic energy and the thermal fluctuations are
\begin{equation}
E_u(k)\sim(\beta g)^{4/5}\varepsilon_\theta^{2/5}k^{-11/5}
\label{eq:esu}
\end{equation}
and
\begin{equation}
E_\theta(k)\sim(\beta g)^{-2/5}\varepsilon_\theta^{4/5}k^{-7/5},
\label{eq:est}
\end{equation}
respectively. Figure \ref{fig:fig12} shows the normalized kinetic-energy and thermal spectra, $E_u(k)/\varepsilon_\theta^{2/5}$ and $E_\theta(k)/\varepsilon_\theta^{4/5}$, obtained at five distinct times during the RT evolution. Here, $E_u(k)$ and $E_\theta(k)$ are calculated by first Fourier transforming of the velocity and temperature fields on one-dimensional horizontal planes and then averaging over different $z$ inside the mixing zone. It is seen that the velocity and temperature spectra have a roll-off rate of $-11/5$ and $-7/5$, respectively, consistent with the BO59 scaling (\ref{eq:esu}) and (\ref{eq:est}). This is particularly evident from the compensated spectra plotted in the insets. In addition, all these spectra collapse well on top of each other in the intermediate range of wavenumbers, corresponding to the scales in the inertial range, and the growth of the integral scale for both velocity and temperature at small wavenumbers is well reproduced by the temporal evolution of the compensated spectra. Taken together, Figs. \ref{fig:fig9}-\ref{fig:fig12} support the BO59 scenario for 2D RT turbulence and the BO59 scaling (\ref{eq:sfu}) and (\ref{eq:sft}) for the cascades of the velocity and temperature fluctuations.

%This type of scaling has been observed in previous numerical studies from the aspect of structure functions \cite{celani2006prl, biferale2010pof} and the intermittency for higher moments of Eq. (\ref{eq:sfu}) and (\ref{eq:sft}) has also been discussed in great detail \cite{celani2006prl, biferale2010pof}. We thus don't want to discuss again structure functions in this paper. Instead,

\subsection{Spatial and temporal intermittency}

To reveal the spatial and temporal intermittency effects of RT turbulence, we extend the dimensional predictions (\ref{eq:sfu}) and (\ref{eq:sft}) to higher-order moments of the fluctuating fields. Therefore, $p$th-order structure functions of velocity and temperature fluctuations are expected to follow, respectively,
\begin{equation}
\label{eq:sfu_p}
S_p(r,t)\equiv\langle|\delta u_r(t)|^p\rangle_V\simeq v_{rms}(t)^p(\frac{r}{h(t)})^{\zeta^r_p}\sim r^{\zeta^r_p}t^{\zeta^t_p}
\end{equation}
and
\begin{equation}
\label{eq:sft_p}
R_p(r,t)\equiv\langle|\delta\theta_r(t)|^p\rangle_V\simeq \Theta_0^p(\frac{r}{h(t)})^{\xi^r_p}\sim r^{\xi^r_p}t^{\xi^t_p}.
\end{equation}
Here, we mainly consider longitudinal velocity and temperature structure functions over horizontal separations. $S_p(r)$ and $R_p(r)$ over vertical separations are also calculated and the similar results are obtained. Mean-field theory predicts $\zeta^r_p=3p/5$, $\zeta^t_p=-p/5$, $\xi^r_p=p/5$, and $\xi^t_p=-2p/5$, while the intermittency effects may result in a deviation with respect to these linear behaviors.

%Here, we consider the isotropic contribution of the statistics, by averaging over all directions of separations.

\begin{figure}[t]
  \centering
 \includegraphics[width=0.49\columnwidth]{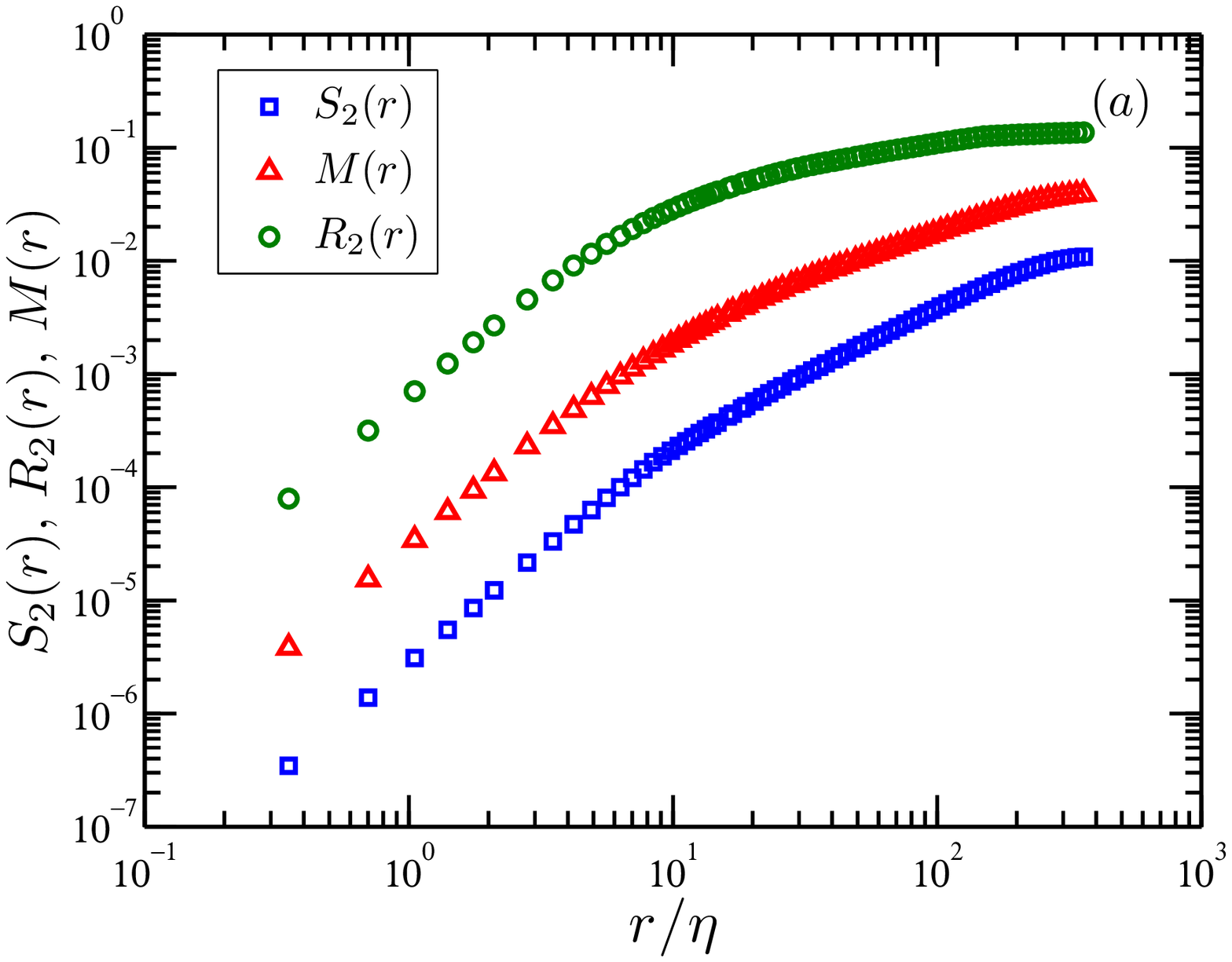}
 \includegraphics[width=0.49\columnwidth]{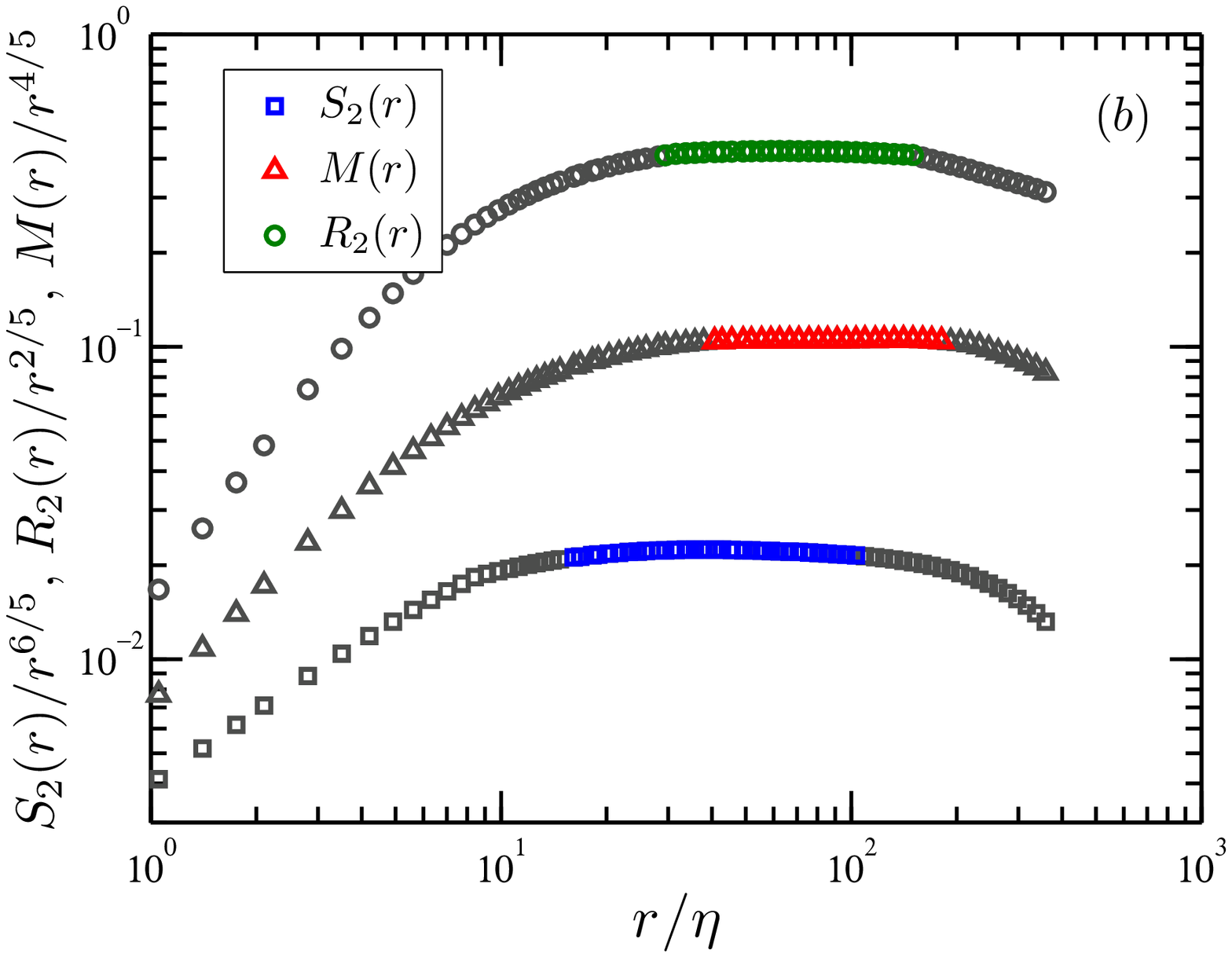}
  \caption{(Color online) (\emph{a}) Second-order velocity and temperature structure functions, $S_2(r)$ and $R_2(r)$, and mixed velocity-temperature structure function $M(r)$ obtained at time $t/\tau=3$. (\emph{b}) The corresponding structure functions compensated with the BO59 scaling. Data inside and outside of the scaling range are represented by different colors. The data in (\emph{b}) have been shifted upwards for clarity.}
  \label{fig:fig13}
\end{figure}

Before the discussion of intermittency, we first look at the second-order statistics, which are most-studied quantities in buoyancy-driven turbulence \cite{lx2010arfm,celani2006prl,boffetta2010pof}. Figure \ref{fig:fig13}(\emph{a}) plots the second-order velocity and temperature structure functions $S_2(r)$ and $R_2(r)$, computed at a late stage $t/\tau=3$ of the self-similar regime. In the figure, we also plot the mixed velocity-temperature structure function,
\begin{equation}
\label{eq:mixsf}
M(r,t)\equiv\langle|\delta\theta_r(t)\delta u_r(t)|\rangle_V,
\end{equation}
which is expected to scale as $M(r)\sim r^{4/5}$ in the BO59 scenario. It is seen that all the computed structure functions display a range of linear scaling. To see this more clearly, we replot them in Fig. \ref{fig:fig13}(\emph{b}) in compensated form in such a way that the expected behavior in the inertial range would be given by a constant, respectively, $S_2(r)/r^{6/5}$, $R_2(r)/r^{2/5}$, and $M(r)/r^{4/5}$ vs $r/\eta$. A plateau is observed for all the three quantities, i.e. $16\lesssim r/\eta\lesssim103$ for $S_2(r)$, $30\lesssim r/\eta\lesssim150$ for $R_2(r)$, and $40\lesssim r/\eta\lesssim180$ for $M(r)$. Note that $R_2(r)$ and $M(r)$ both have a range of scaling which extends to larger scales, which may be due to the large-scale temperature structures, like plumes or spikes, which have strong correlations with vertical velocity. We further note that this feature is qualitatively consistent with those observed in 3D cases \cite{boffetta2010pof}.

\begin{figure}[t]
  \centering
 \includegraphics[width=0.49\columnwidth]{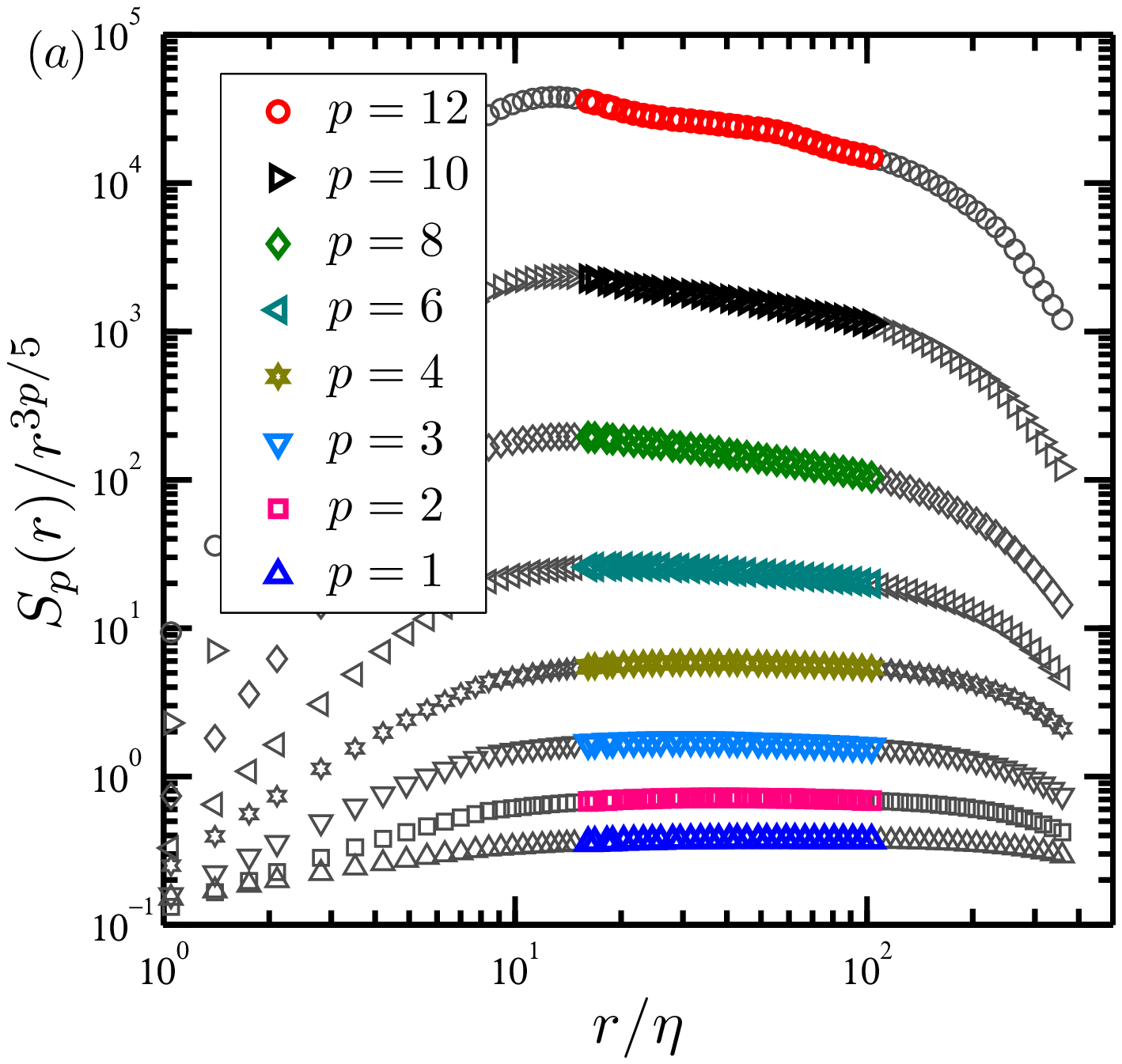}
 \includegraphics[width=0.49\columnwidth]{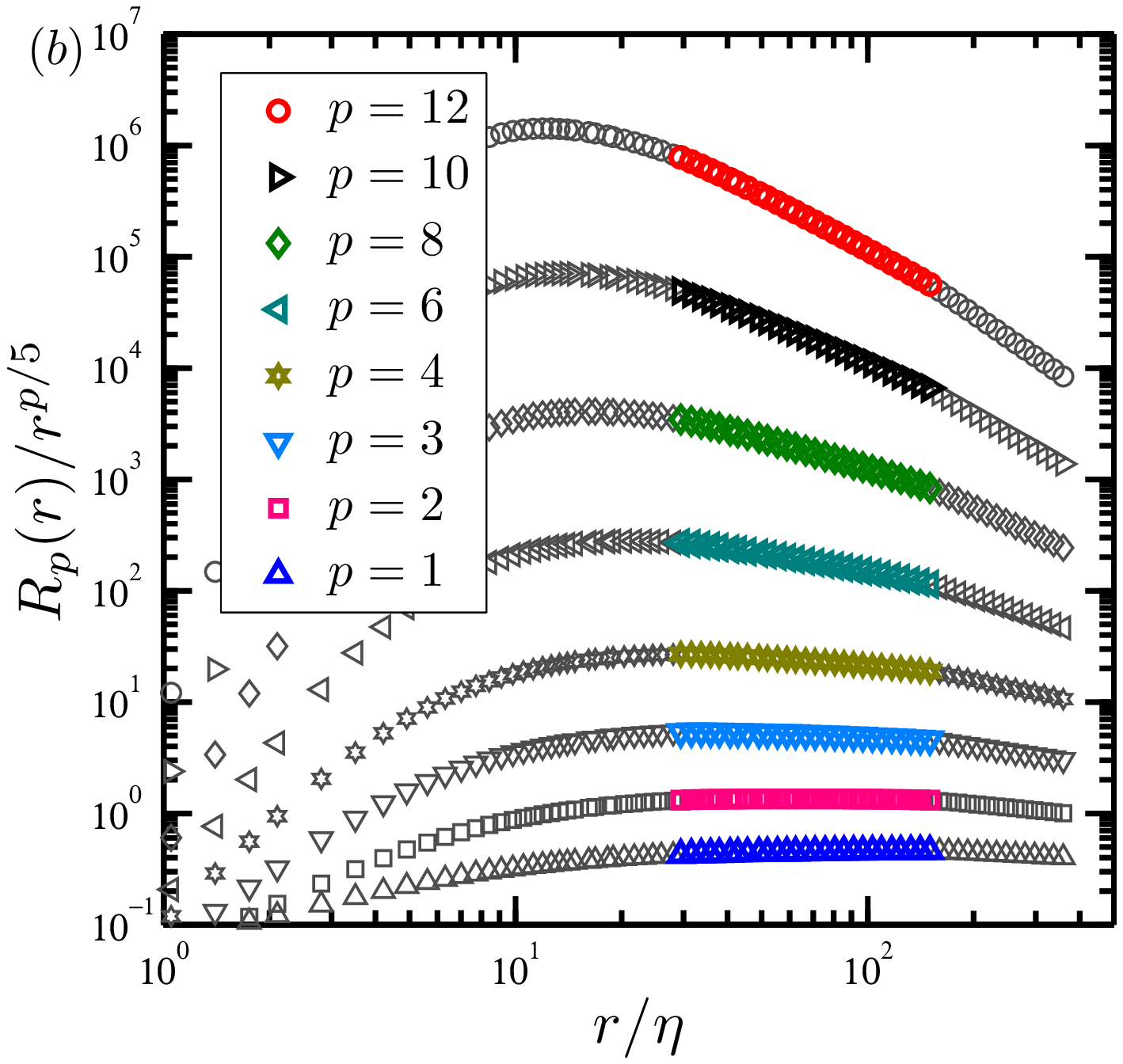}
  \caption{(Color online) Plots of the $p$th-order structure functions of velocity (\emph{a}) and temperature (\emph{b}) compensated with the BO59 scaling. Data inside and outside of the scaling range are represented by different colors. For clarity, the data have been shifted upwards for both velocity and temperature. The data are obtained at time $t/\tau=3$.}
  \label{fig:fig14}
\end{figure}

\begin{figure}[t]
  \centering
 \includegraphics[width=0.485\columnwidth]{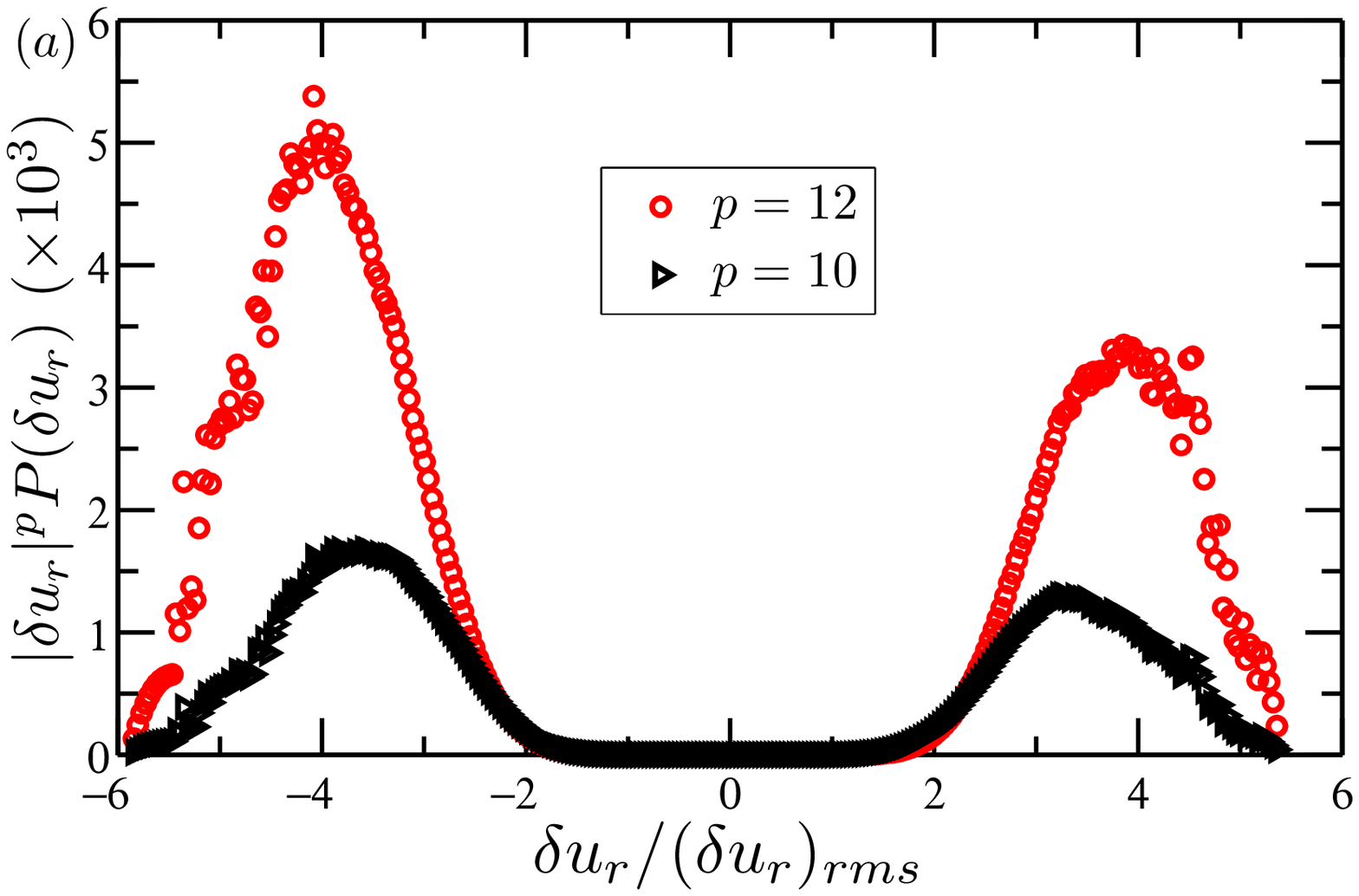}
 \includegraphics[width=0.495\columnwidth]{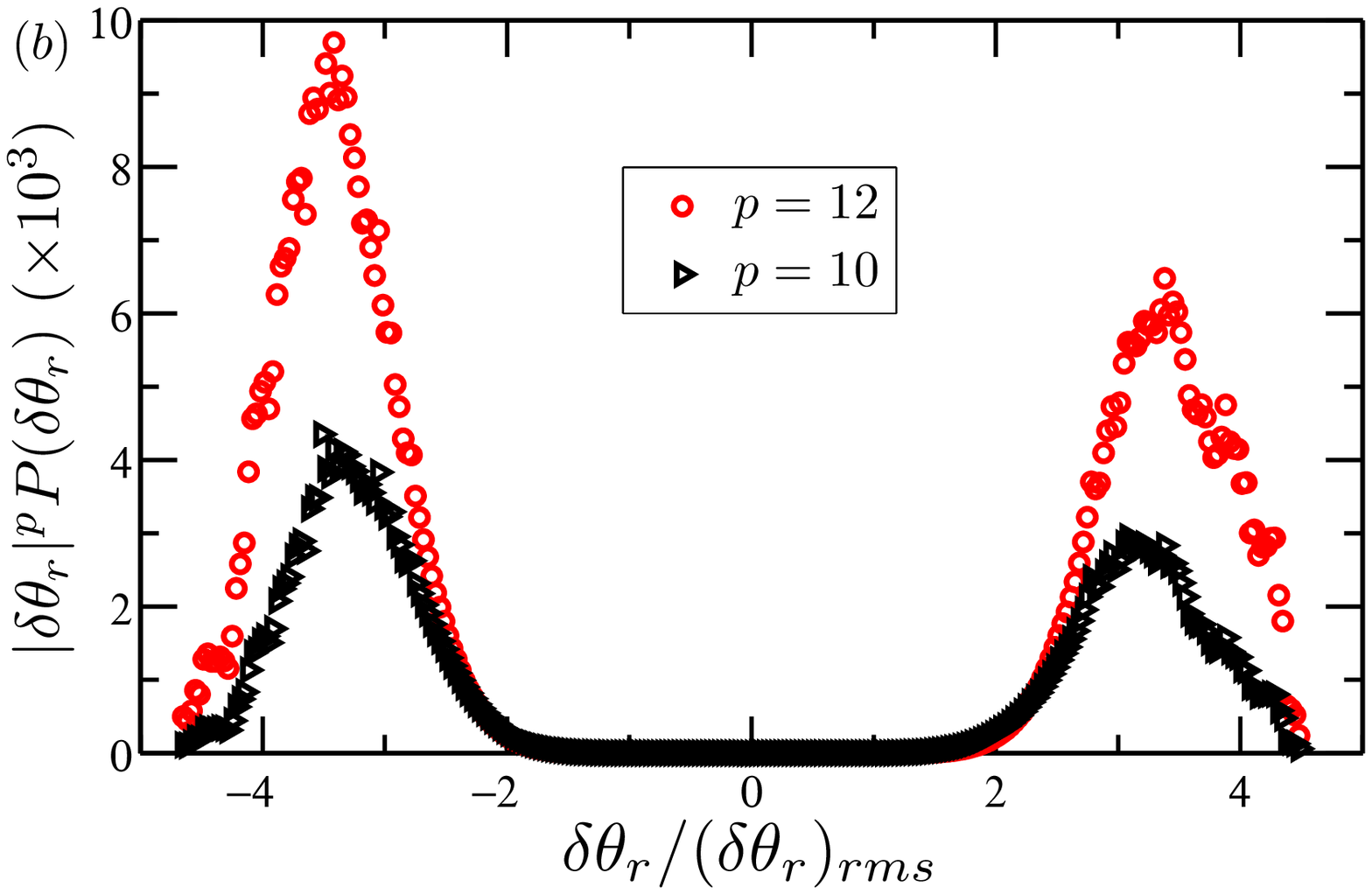}
  \caption{(Color online) Integral kernels of velocity (\emph{a}) and temperature (\emph{b}) structure functions at the lower end of the inertial range for $p=10$ and 12. The vertical scale for $p=10$ has been enlarged by 5 times for both velocity and temperature. The data are obtained at time $t/\tau=3$.}
  \label{fig:fig15}
\end{figure}

Figure \ref{fig:fig14} plots in log-log scale $S_p(r)/r^{3p/5}$ and $R_p(r)/r^{p/5}$ as functions of $r$ obtained at $t/\tau=3$ for $p=1$, 2, 3, 4, 6, 8, 10, and 12 (from bottom to top). Data inside and outside of the scaling range are represented by different colors. Figure \ref{fig:fig14} shows that the structure functions for both velocity and temperature exhibit good scaling. The compensated plot shows the quality of the structure functions and their progressive deviation from the BO59 prediction with increasing $p$. To show the level of convergence of these structure functions, we examine their integration kernels, $|\delta u_r|^pP(\delta u_r)$ of $S_p(r)$ and $|\delta\theta_r|^pP(\delta\theta_r)$ of $R_p(r)$, which are shown in Fig. \ref{fig:fig15} for $p=10$ and 12. Here, $P(\delta u_r)$ and $P(\delta\theta_r)$ are, respectively, probability density functions (PDFs) of $\delta u_r$ and $\delta\theta_r$. The figure shows that $S_p(r)$ and $R_p(r)$ both exhibit very good convergence even for the highest order $p=12$. The figure also shows that the integration kernels are very asymmetric for both velocity and temperature, a signature of persistence of cliff-ramp-like structures of the velocity and temperature fields, like fronts of plumes/spikes \cite{zhou2002prl,zhou2008pre,zhou2011jfm,zhou2011pof}.

\begin{figure}[t]
  \centering
 \includegraphics[width=0.48\columnwidth]{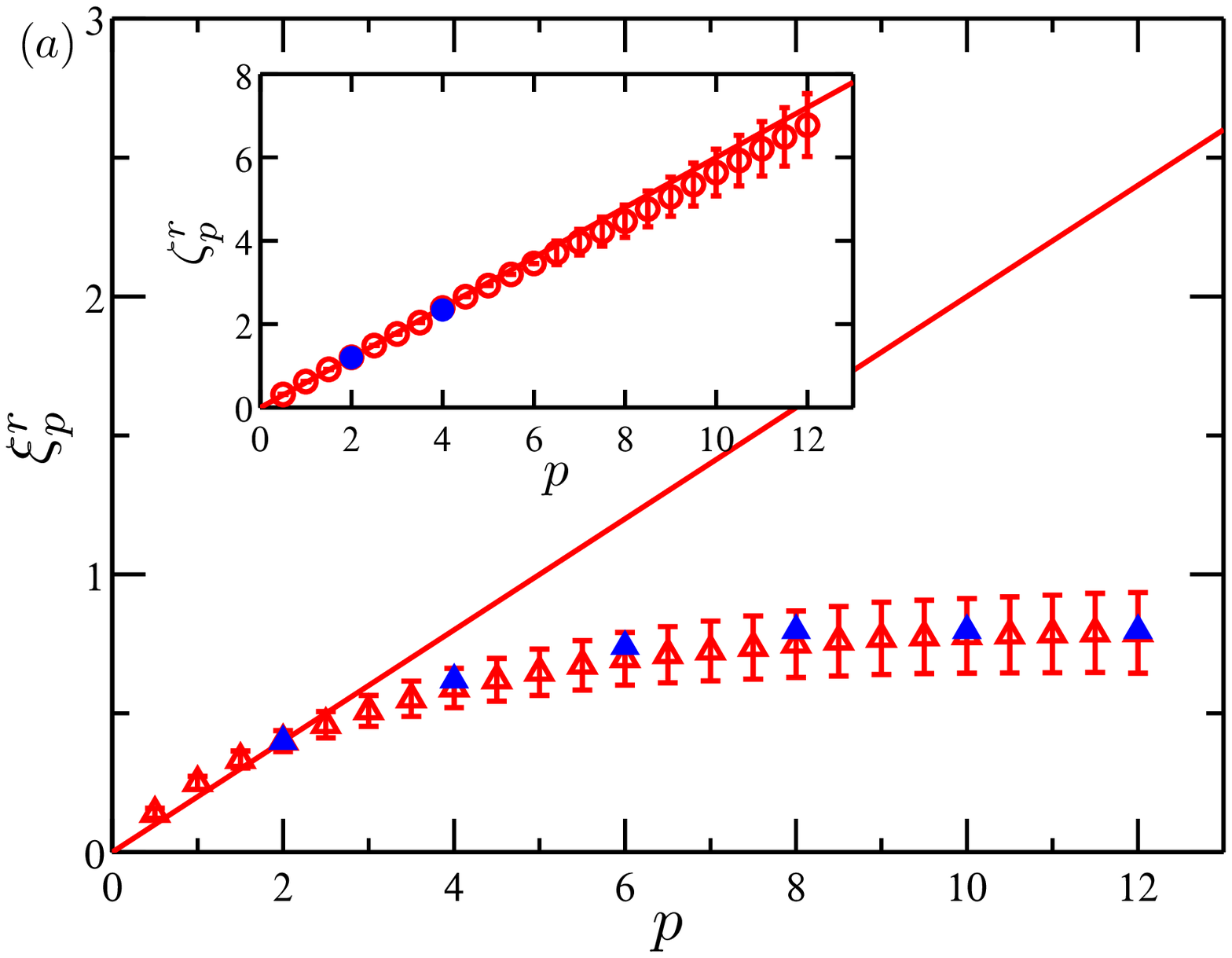}
 \includegraphics[width=0.49\columnwidth]{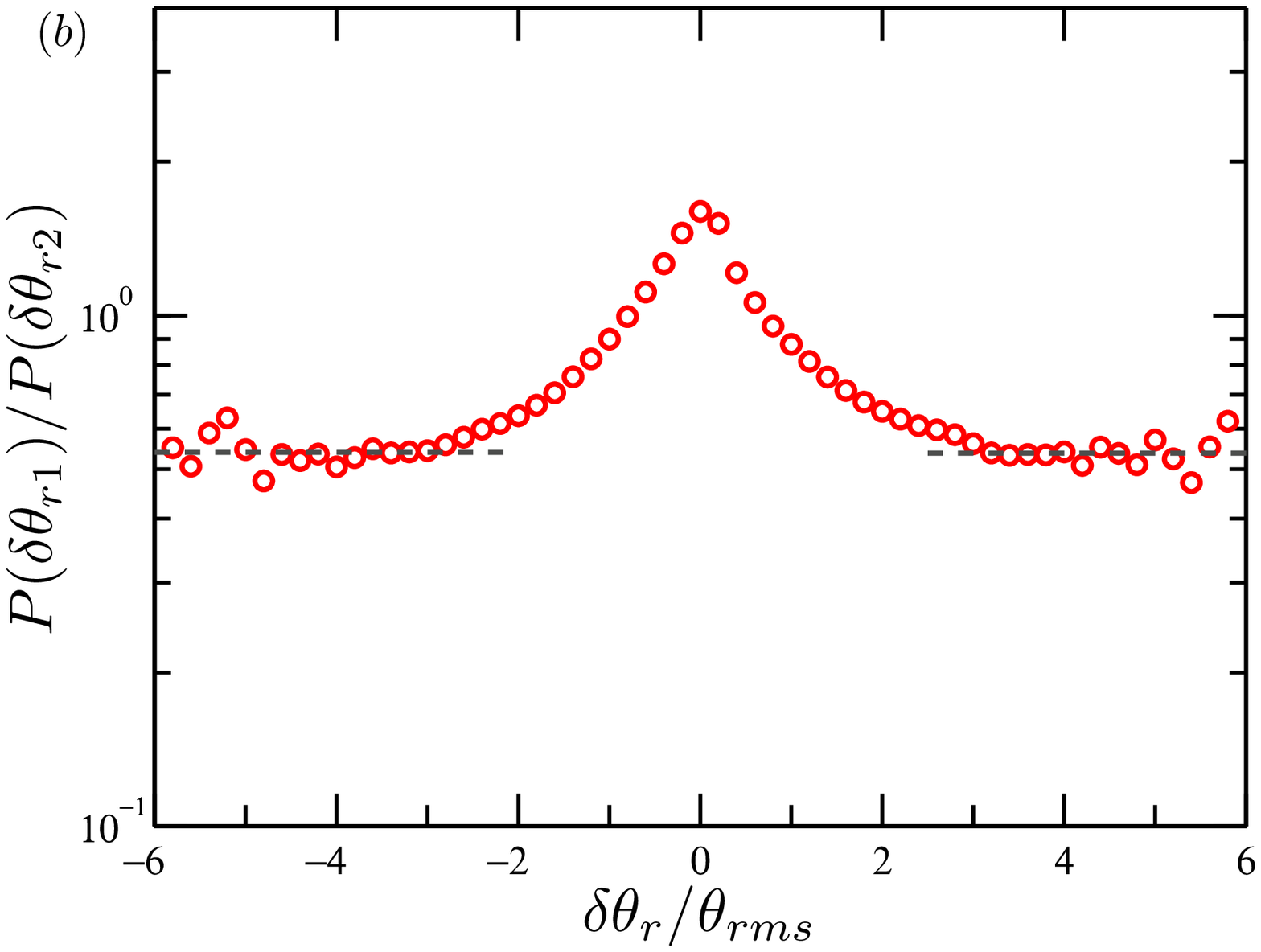}
  \caption{(Color online) (\emph{a}) Temperature structure function exponents, $\xi^r_p$. Inset: Velocity structure function exponents, $\zeta^r_p$. The straight lines are the dimensional predictions, $p/5$ for temperature and $3p/5$ for velocity. Red open symbols are obtained from the present simulations at time $t/\tau=3$. Error bars indicate the variations of $\zeta^r_p$ and $\xi^r_p$ among different realizations. Blue filled symbols are taken from a RB simulation at $Ra=10^7$ (Ref.\cite{celani2001pof}). (\emph{b}) Ratio of two PDFs, at different inertial scales, $r_1=43\eta$ and $r_2=113\eta$, for temperature differences. The horizontal dashed lines mark the constant ratio of the PDF tails.}
  \label{fig:fig16}
\end{figure}

The scaling exponents $\xi^r_p$ of temperature structure functions of orders up to $p=12$ are plotted in Fig. \ref{fig:fig16}(\emph{a}) and the velocity exponents $\zeta^r_p$ are shown in the inset. Different realizations of 2D RT turbulence result in fluctuations of scaling exponents which account for the error bars shown in Fig. \ref{fig:fig16}. One sees that the velocity statistics is quite close to the BO59 dimensional prediction, i.e. $\zeta^r_p=3p/5$. One also sees that $\zeta^r_p$ is slightly, but systematically smaller than the linear scaling for $p\gtrsim6$, suggesting the possible presence of some degree of intermittency. Indeed, a small intermittent correction to the 2D BO59 scaling for the velocity field can be detected when concerning gradients evolution \cite{biferale2010pof}.

On the other hand, the values of $\xi^r_p$ are seen to strongly depart from the linear law $p/5$ and the gap increases with the order, a usual signature of inertial range intermittency. An interesting feature worthy of note is that the temperature exponents $\xi^r_p$ are observed to increase extremely slowly with order $p$, suggesting a saturation for $p$ around 10 at a value
\begin{equation}
\label{eq:saturation}
\xi^r_{\infty}=0.78\pm0.15.
\end{equation}
A consequence of this saturation is a constant ratio of the far tails of the PDFs for inertial range separations \cite{celani2001pof,tabeling2001prl,zhou2002prl}. The ratio of two PDFs, for separations $r_1=43\eta$ and $r_2=113\eta$ well into the inertial range, is plotted in Fig. \ref{fig:fig16}(\emph{b}). For temperature increments $|\delta\theta_r|\gtrsim3\theta_{rms}$, this ratio indeed tends as expected towards a constant. This saturation is a feature shared by several scalar turbulent systems, e.g., for passive scalar in Navier-Stokes turbulence \cite{celani2001pof,tabeling2001prl} and for active scalar in buoyancy-driven turbulence \cite{celani2002prl,zhou2002prl}. Especially, the present obtained $\xi^r_{\infty}$ agrees well with the value of 0.8 found in turbulent RB convection \cite{celani2002prl,zhou2002prl}. This prompts us to directly compare scaling exponents with those obtained in the RB system. As shown in Fig. \ref{fig:fig16}(\emph{a}), where blue filled symbols are taken from a RB simulation \cite{celani2001pof} at $Ra=10^7$ and red open symbols are obtained from the present simulations at time $t/\tau=3$, the two sets of exponents coincide well with each other within the error bars. As the RT and RB systems are dominated by the same equations but with different boundary and initial conditions, the present results support the universality scenario in which the set of velocity and active scalar scaling exponents are independent of the details of boundary and initial conditions of the flow.

\begin{figure}[t]
  \centering
 \includegraphics[width=0.49\columnwidth]{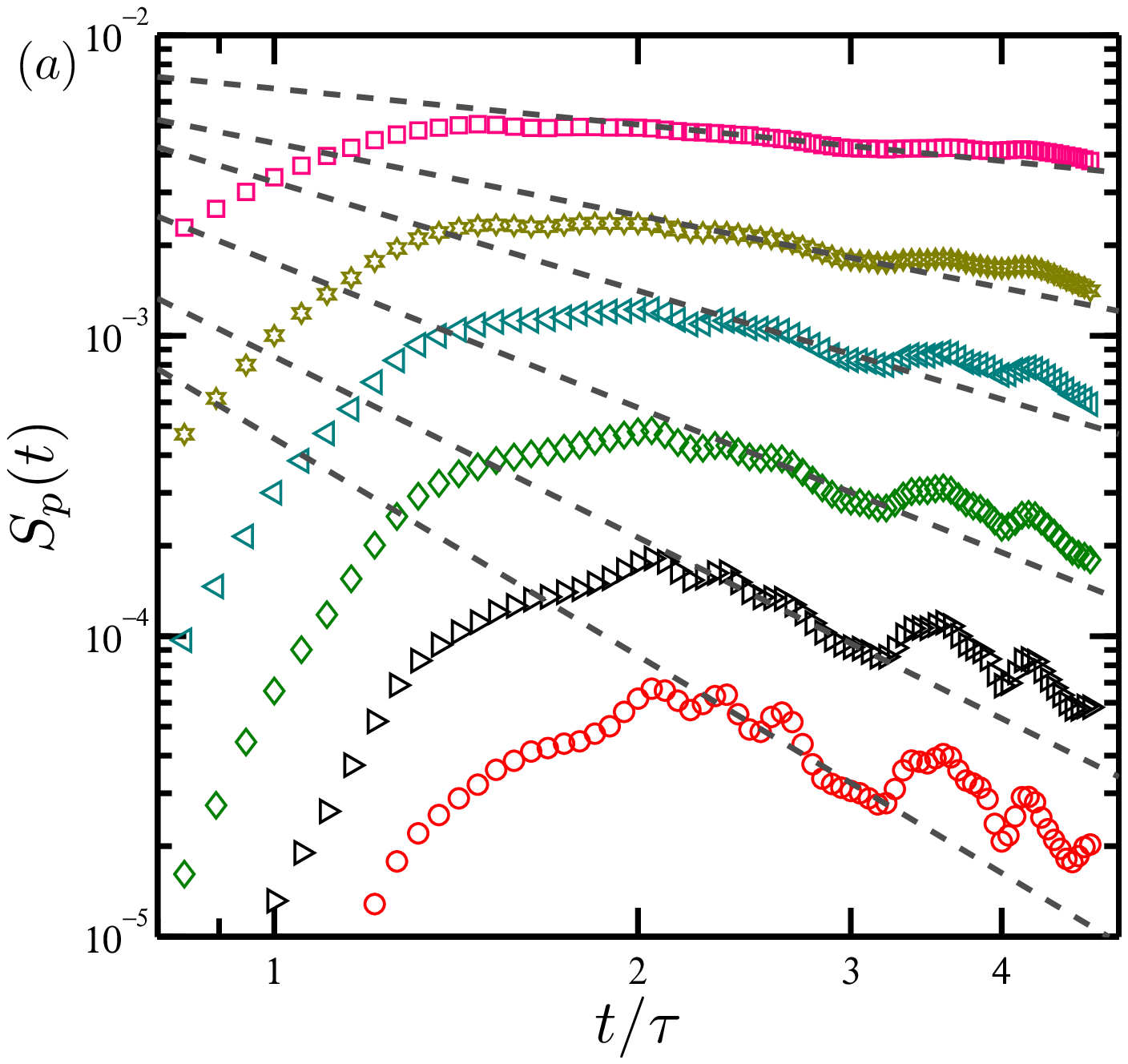}
 \includegraphics[width=0.49\columnwidth]{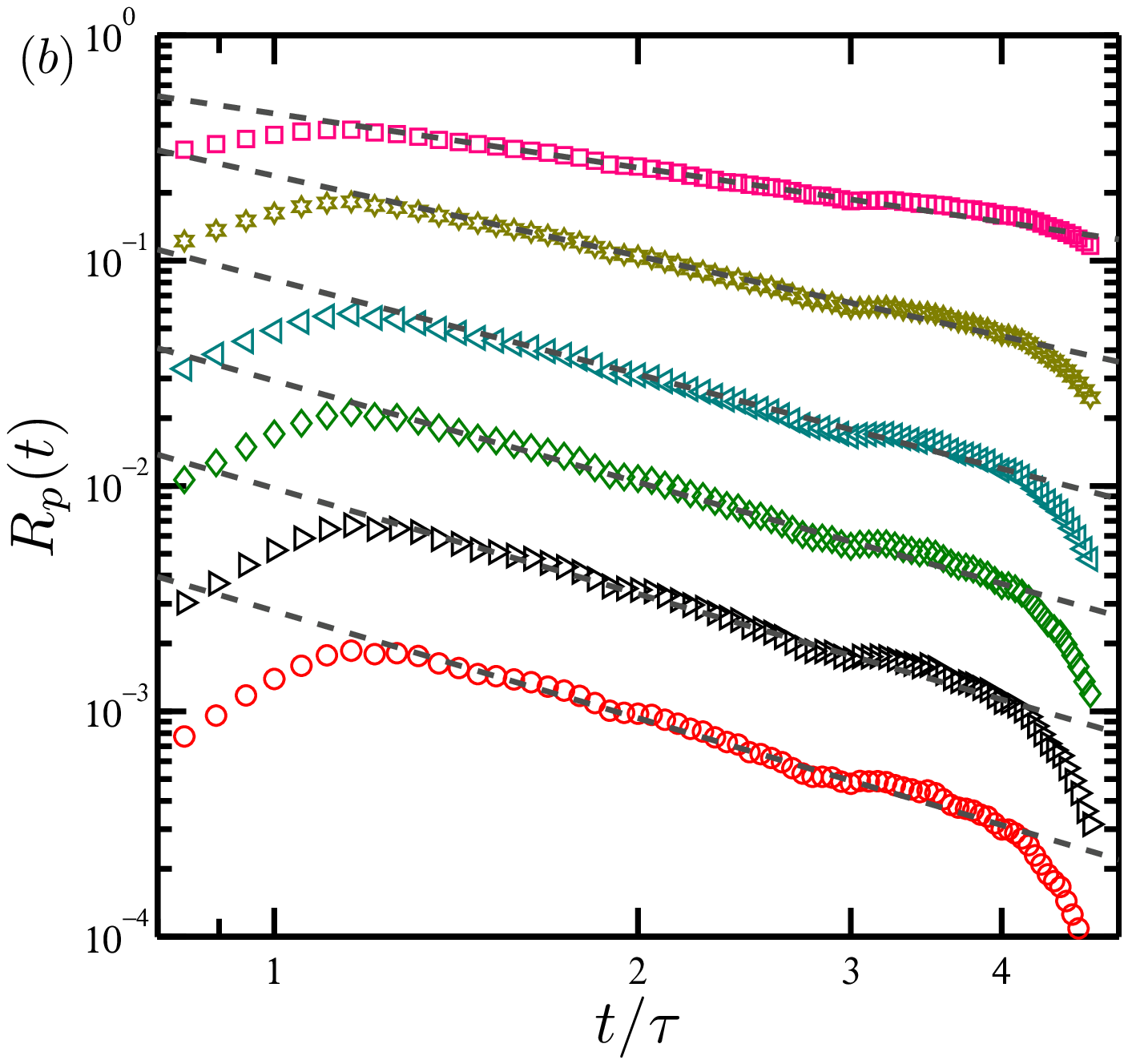}
  \caption{(Color online) Time dependence of velocity (\emph{a}) and temperature (\emph{b}) structure functions, $S_p(t)$ and $R_p(t)$, of order $p=2$ (squares), 4 (hexagrams), 6 (left-triangles), 8 (diamonds), 10 (right-triangles), and 12 (circles) (from top to bottom). The data are obtained at the scale $r=43\eta$ within the inertial ranges for both velocity and temperature. The dashed lines correspond to the dimensional scaling $\zeta^t_p=-p/5$ for velocity and to the intermittency-corrected scaling $\xi^t_p=-2\xi^r_p$ for temperature with $\xi^r_p$ given by Fig. \ref{fig:fig16}(\emph{a}).}
  \label{fig:fig17}
\end{figure}

We finally examine the temporal evolution of structure functions, $S_p(t)$ and $R_p(t)$. From Eq. (\ref{eq:sfu_p}) and taking into account the temporal behaviors of global quantities $v_{rms}(t)$ and $h(t)$ [i.e. Eqs. (\ref{eq:ht}) and (\ref{eq:urms})], $S_p(t)$ is expected to follow the temporal scaling $S_p(t)\sim t^{\zeta^t_p}$ with $\zeta^t_p=p-2\zeta^r_p$. With the BO59 velocity scaling one immediately has $\zeta^t_p=-p/5$. Figure \ref{fig:fig17}(\emph{a}) shows $S_p(t)$ as a function of $t/\tau$ computed at the scale $r=43\eta$ that belongs to the inertial scales. The dashed lines in the figure indicate the dimensional scaling $\zeta^t_p=-p/5$. One sees that although the data show some degree of fluctuations, especially for the orders larger than 8, the dimensional predication can roughly fit the data within the range $2\lesssim t/\tau\lesssim3$. For the temperature field, Eq. (\ref{eq:sft_p}) gives $\xi^t_p=-2\xi^r_p$ with $\xi^r_p$ being the anomalous scaling exponents for temperature. The temporal evolution of $R_p(t)$ is shown in Fig. \ref{fig:fig17}(b) and the dashed lines in the figure indicate the intermittency-corrected scaling $\xi^t_p=-2\xi^r_p$ obtained from the spatial exponents $\xi^r_p$ of Fig. \ref{fig:fig16}(\emph{a}). It is seen that $R_p(t)$ exhibits a much smoother and wider scaling range than $S_p(t)$ and the intermittency-corrected prediction can well describe the data within the range $1\lesssim t/\tau\lesssim3$.

\section{Conclusion}

In conclusion, we have carried out a careful investigation of the temporal evolution and the scaling behavior of RT turbulence in two dimensions, by means of high-resolution direct numerical simulations. We present results from an ensemble of 100 independent realizations performed at unit Prandtl number and small Atwood number with a spatial resolution of $2048\times8193$ grid points and Rayleigh number up to $Ra\sim10^{11}$. The major findings can be summarized as follows:
% Our main objective is to deepen the previous numerical studies \cite{celani2006prl,biferale2010pof} and test the validation of the theoretical predictions of the Chertkov's 2D model \cite{chertkov2003prl}.
\begin{enumerate}
  \item For small-scale turbulent properties, a force balance is found between the buoyancy force and the inertial force at all scales below the integral length scale and thus validate of the basic force-balance assumption of the BO59 scenario in 2D RT turbulence. It is further found that the Kolmogorov dissipation scale $\eta(t)\sim t^{1/8}$, the kinetic-energy dissipation rate $\varepsilon_u(t)\sim t^{-1/2}$, and the thermal dissipation rate $\varepsilon_{\theta}(t)\sim t^{-1}$. All these scaling properties are in excellent agreement with the theoretical predictions of the Chertkov model for the 2D case \cite{chertkov2003prl}.

  \item The statistics of velocity differences is nearly self-similar in 2D RT turbulence even if small intermittent corrections cannot be ruled out, especially for the orders larger than 6, while the scaling exponents $\xi^r_p$ of temperature structure functions are found to strongly deviate from the dimensional prediction. Furthermore, $\xi^r_p$ tends to saturate at high orders to $\xi^r_{\infty}=0.78\pm0.15$, a signature of persistence of plumes/spikes fronts even at very small scales \cite{celani2001pof}. The value of $\xi^r_{\infty}$ and the order at which saturation occurs are compatible with those of turbulent RB convection \cite{celani2002prl}, supporting the scenario of universality of buoyancy-driven turbulence with respect to the different boundary conditions characterizing the RT and RB systems.

  \item For the statistics of global quantities, the 2D large-scale vortices in the mixing zone grow much faster than their 3D counterparts and the 2D RT turbulence possesses a relatively lower degree of anisotropy compared to the 3D case.

  \item In two dimensions the kinetic-energy dissipation rate becomes neglectable in the turbulence regime and hence almost all the energy injected into the flow contributes to the growth of the large-scale flow.
\end{enumerate}

%\label{}
%\subsubsection{}

% If in two-column mode, this environment will change to single-column format so that long equations can be displayed.
% Use only when necessary.
%\begin{widetext}
%$$\mbox{put long equation here}$$
%\end{widetext}

% Figures should be put into the text as floats.
% Use the graphics or graphicx packages (distributed with LaTeX2e).
% See the LaTeX Graphics Companion by Michel Goosens, Sebastian Rahtz, and Frank Mittelbach for examples.
%
% Here is an example of the general form of a figure:
% Fill in the caption in the braces of the \caption{} command.
% Put the label that you will use with \ref{} command in the braces of the \label{} command.
%
% \begin{figure}
% \includegraphics{}%
% \caption{\label{}}%
% \end{figure}

% Tables may be be put in the text as floats.
% Here is an example of the general form of a table:
% Fill in the caption in the braces of the \caption{} command. Put the label
% that you will use with \ref{} command in the braces of the \label{} command.
% Insert the column specifiers (l, r, c, d, etc.) in the empty braces of the
% \begin{tabular}{} command.
%
% \begin{table}
% \caption{\label{} }
% \begin{tabular}{}
% \end{tabular}
% \end{table}

\begin{acknowledgments}
This work was supported by the Natural Science Foundation of China (NSFC) under Grant Nos. 11222222, 11272196, and 11161160554 and Innovation Program of Shanghai Municipal Education Commission under Grant Nos. 13YZ008 and 11ZZ87.
\end{acknowledgments}

% Create the reference section using BibTeX:
%\bibliography{RTT} % Compile the BIBTEX file on your computer to generate the BBL file. Then copy and paste the BBL file contents below.
%\begin{bibliography)
%\end{bibliographiy)
%merlin.mbs aipnum4-1.bst 2010-07-25 4.21a (PWD, AO, DPC) hacked
%Control: key (0)
%Control: author (8) initials jnrlst
%Control: editor formatted (1) identically to author
%Control: production of article title (0) allowed
%Control: page (1) range
%Control: year (1) truncated
%Control: production of eprint (0) enabled
%

\end{document}